\documentclass[conference]{IEEEtran}
\usepackage[hyphens]{url}
\usepackage{hyperref}
\hypersetup{breaklinks=true, colorlinks,allcolors=blue}
\usepackage[backend=bibtex, style=ieee]{biblatex}

\IEEEoverridecommandlockouts
\usepackage{amsmath,amssymb,amsfonts}
\usepackage{algorithmic}
\usepackage{graphicx}
\usepackage{textcomp}
\usepackage{xcolor}  
\usepackage{balance}
\usepackage{mathtools}  
\usepackage{booktabs}  
\usepackage{multirow}  
\usepackage{orcidlink}  
\usepackage{kantlipsum} 
\usepackage{float}
\usepackage{stfloats}
\usepackage{enumitem}
\usepackage{comment}

\def\BibTeX{{\rm B\kern-.05em{\sc i\kern-.025em b}\kern-.08em
    T\kern-.1667em\lower.7ex\hbox{E}\kern-.125emX}}

\begin{document}

\title{Quantifying Return on Security Controls in LLM Systems}
\author{\IEEEauthorblockN{Richard Helder Moulton\,\orcidlink{0009-0005-5568-1012}, Austin O'Brien\,\orcidlink{0009-0004-8479-6612}, John D. Hastings\,\orcidlink{0000-0003-0871-3622}}
\IEEEauthorblockA{\textit{The Beacom College of Computer \& Cyber Sciences} \\
\textit{Dakota State University }\\
Madison, SD, USA \\
rich.moulton@trojans.dsu.edu, \{austin.obrien,john.hastings\}@dsu.edu}

}

\maketitle

\begin{abstract}
Although large language models (LLMs) are increasingly used in security-critical workflows, practitioners lack quantitative guidance on which safeguards are worth deploying. This paper introduces a decision-oriented framework and reproducible methodology that together quantify residual risk, convert adversarial probe outcomes into financial risk estimates and return-on-control (RoC) metrics, and enable monetary comparison of layered defenses for LLM-based systems. A retrieval-augmented generation (RAG) service is instantiated using the DeepSeek-R1 model over a corpus containing synthetic personally identifiable information (PII), and subjected to automated attacks with Garak across five vulnerability classes: PII leakage, latent context injection, prompt injection, adversarial attack generation, and divergence. For each (vulnerability, control) pair, attack success probabilities are estimated via Laplace's Rule of Succession and combined with loss triangle distributions, calibrated from public breach-cost data, in 10,000-run Monte Carlo simulations to produce loss exceedance curves and expected losses. Three widely used mitigations, attribute-based access control (ABAC); named entity recognition (NER) redaction using Microsoft Presidio; and NeMo Guardrails, are then compared to a baseline RAG configuration. The baseline system exhibits very high attack success rates ($\geq$ 0.98 for PII, latent injection, and prompt injection), yielding a total simulated expected loss of \$313k per attack scenario. ABAC collapses success probabilities for PII and prompt-related attacks to near zero and reduces the total expected loss by $\approx$ 94\%, achieving an RoC of 9.83. NER redaction likewise eliminates PII leakage and attains an RoC of 5.97, while NeMo Guardrails provides only marginal benefit (RoC of 0.05).
\end{abstract}



\section{Introduction}
\label{sec:Introduction}
Large language models (LLMs) are increasingly deployed across diverse fields, including healthcare, finance, education, law, medicine, and cybersecurity \cite{feretzakis2024trustworthy,lai2024llmlaw,zhou2023survey,zhang2025llms}. Their widespread adoption is driven by their capabilities, such as natural language understanding, contextual reasoning, text generation, summarization, translation, and code synthesis~\cite{brown2020language,bommasani2022opportunities,chen2021evaluatingllms}. These models facilitate automation in customer service, legal document drafting, medical report generation, and software development~\cite{bommasani2022opportunities,lai2024llmlaw,zhou2023survey}. Furthermore, their ability to perform few-shot and zero-shot learning enables adaptation to new tasks with minimal domain-specific training, accelerating their integration into various industries~\cite{brown2020language,bubeck2023sparks}. 

Despite their advantages, LLMs also exhibit important security and trustworthiness challenges that limit their safe and effective use. Studies have identified vulnerabilities including adversarial attacks, prompt injection, data leakage, and model bias, raising concerns about their robustness and real world applicability \cite{RN245, RN248, RN249, jain2023baseline, mozes2023use}. In high stakes contexts, these weaknesses can result in failures on sentiment related tasks, unintended disclosure of sensitive information, susceptibility to jailbreak and prompt injection attempts, hallucinated or fabricated content, and biased or otherwise inappropriate outputs with harmful societal consequences \cite{hu2024firm, chen2023can}.

\subsection{Statement of the Problem}
In their current form, LLMs lack consistent safeguards capable of preventing these vulnerabilities in the face of adaptive, sophisticated attacks~\cite{zou2023universal,jain2023baseline,RN248}.
Addressing these risks requires reliable, targeted, and computationally affordable mitigation strategies.

A variety of mitigation strategies have been proposed including Attribute-Based Access Control (ABAC)~\cite{hu2013guide} as a higher-level system control, Named Entity Recognition (NER) ~\cite{feretzakis2024trustworthy} as a component of content filtering or detection, and differential privacy~\cite{abadi2016deep} as a technique to limit the leakage of sensitive information during training or generation. Other proposed defenses include perplexity filtering~\cite{wenzek2020ccnet}, input preprocessing (e.g., paraphrasing and retokenization)~\cite{jain2023baseline}, adversarial training~\cite{miyato2017adversarial}, the use of auxiliary classifier models (``detectors'')~\cite{gehman2020real}, red-teaming frameworks~\cite{liang2023holistic}, and intention analysis~\cite{zhang2025intention}. 
However, a critical gap remains in understanding the real-world effectiveness and computational costs associated with deploying or not deploying these safeguards, especially against sophisticated and adaptive attacks. 

\subsection{Objectives of the Project}
To address this gap, 
this study aims to 
systematically evaluate, quantify, and analyze the effectiveness and feasibility of various safeguards designed to mitigate known vulnerabilities in LLMs. Specifically, it will: 

\begin{itemize}
    \item Assess the effectiveness of known LLM vulnerability safeguards by applying them to a model with well-documented security flaws \cite{schultz2025deepseek,wired2025deepseek}. This includes evaluating how well different mitigation techniques prevent unintended information leakage, resist adversarial manipulations such as prompt injections and jailbreaks, and reduce biases, hallucinations, and other undesirable behaviors.
    
    \item Measure and analyze the costs of implementing or eschewing these safeguards, considering factors such as cost of implementation and cost of breach. By quantifying the resource demands of each approach, this study will provide insights into the trade-offs between security risk, costs, and practical deployment feasibility.
    
    \item Compare multiple mitigation techniques across different vulnerability categories to determine their strengths, weaknesses, and suitability for real-world applications. This comparative analysis will highlight the most effective and cost-efficient strategies while identifying areas where further optimization or refinement is needed.
    
    \item Evaluate the adaptability and resilience of safeguards against evolving adversarial techniques, including automated jailbreak strategies, adaptive attacks, and sophisticated prompt engineering methods. By testing these defenses under varying conditions, this study will help establish their robustness and reliability.
    
\end{itemize}

\subsection{Significance of the Study}
As LLMs become increasingly embedded in sensitive domains, ensuring their trustworthiness and security is critical. Although a wide range of mitigation strategies exist, including prompt filtering; access control; and privacy-preserving training, current evaluations focus primarily on qualitative security outcomes or functional correctness with limited attention to modeling financial or operational risk \cite{mazeika2024harmbenchstandardizedevaluationframework, brokman2024insights, zhu2023promptbench}. As a result, the cost effectiveness of mitigation strategies remains underexplored. This research addresses that need by delivering a comprehensive assessment of multiple safeguards under controlled, adversarial conditions, quantifying both the security benefits and economic costs. 
The findings will help developers, security practitioners, and policy-makers make informed decisions about which protections are most viable for real-world use, enabling a more balanced approach to trustworthy AI that considers not only security but also scalability, usability, and resource constraints.
  
The main contributions of this work are as follows:

\begin{enumerate}
\item A reproducible methodology that integrates probabilistic risk modeling, Monte Carlo simulation, and cost-benefit analysis to evaluate LLM security safeguards.
\item An empirical comparison of three security controls (ABAC, NER, and NeMo Guardrails) against five representative vulnerability classes (personally identifiable information (PII) leakage, latent context injection, adversarial attack generation, direct prompt injection, and divergence).
\item Decision-support metrics, including Return on Controls (RoC), that quantify the reduction in expected loss per unit cost, supporting practical security investment decisions.
\end{enumerate}

The remainder of the paper is organized as follows. Section \ref{sec:related} presents related work; Section \ref{sec:methodology} details the theoretical framework and methodology; Section \ref{sec:results} reports results and discussion; Section \ref{sec:future} outlines future work; and Section \ref{sec:conclusion} concludes the paper.

\section{Related Work}
\label{sec:related}
\subsection{Taxonomies and Attack Vectors}

A growing body of research explores the risks LLMs pose when exposed to adversarial behavior. Cui et al.~\cite{RN245} proposed a comprehensive taxonomy of vulnerabilities organized across four system modules: input, language model, toolchain, and output. This framework provides a modular lens for analyzing threats and has informed subsequent evaluations of LLM behavior under attack.

Zhang et al.~\cite{zhang2025llms} surveyed the use of LLMs in cybersecurity, highlighting both defensive applications and attack surfaces. They emphasized the urgent need for systematic defenses as LLMs are increasingly used in sensitive workflows. Specific attack vectors such as prompt injection, suffix-based jailbreaks, and adversarial prompting have been studied in depth. 

Zou et al.~\cite{zou2023universal} introduced transferable adversarial prompts that can bypass alignment safeguards across models. Carlini et al.~\cite{carlini2023extracting} demonstrated that LLMs may memorize and leak training data, raising serious concerns for privacy and data protection.

Beyond suffix-based and universal jailbreaks, targeted jailbreaking methods and datasets (e.g., \cite{guo2024cold,jin2024jailbreakzoo}) further indicate that carefully crafted prompts can consistently elicit restricted content across models and settings. These risks are amplified in agentic configurations that use external tools or APIs, where indirect prompt injection and cross-domain context attacks enlarge the effective attack surface~\cite{RN249}.

\subsection{Benchmarks and Security Evaluation Tooling}
Several benchmarking frameworks have been proposed to evaluate LLM robustness. HarmBench~\cite{mazeika2024harmbenchstandardizedevaluationframework} standardizes the assessment of harmful prompt generation and model refusal behavior. Brokman et al.~\cite{brokman2024insights} compared popular open-source scanners such as Garak and Giskard, identifying inconsistent coverage of known threats. Li et al.~\cite{li2024privacylargelanguagemodels} focused specifically on privacy risks and proposed PrivLM-Bench to measure vulnerability to membership inference and data extraction.
Complementary 
suites such as PromptBench~\cite{zhu2023promptbench} and related adversarial-prompt tooling~\cite{zhou2024robustpromptoptimizationdefending} help probe jailbreakability and robustness, while framework-level scanners like Garak~\cite{derczynski2024garak} support automated, repeatable security probing in practice.

\subsection{Mitigation Strategies and Privacy-Preserving Approaches}
Mitigation strategies have included both preventive and reactive techniques. Feretzakis and Verykios~\cite{feretzakis2024trustworthy} assessed the use of inference-time filtering, Named Entity Recognition, and access control mechanisms (e.g., RBAC and ABAC) to prevent sensitive information disclosure. However, such approaches may be limited in their coverage or scalability. He et al.~\cite{RN249} further analyzed security risks in LLM agents, which combine language models with external tools and APIs, exposing more complex and less understood attack surfaces.

Privacy-preserving methods complement inference-time controls: differentially private training and tooling (e.g., TensorFlow Privacy)~\cite{papernot2019TensorFlow} reduce the risk of memorization and membership inference; federated and decentralized training paradigms discussed in privacy surveys~\cite{li2024privacylargelanguagemodels} mitigate centralized data exposure; and parameter-efficient adaptation (e.g., LoRA/QLoRA) can limit retraining footprint and reduce inadvertent regressions introduced during fine-tuning~\cite{feffer2024red}. Rule-based and policy-driven guardrail systems (e.g., NeMo Guardrails)~\cite{RN246} are increasingly used to constrain outputs at inference, though their effectiveness depends greatly on configuration and integration depth.

\subsection{Offensive Uses and Operational Threats}
Recent work has explored how LLMs can be weaponized for offensive operations. Mozes et al.~\cite{mozes2023use} cataloged illicit uses such as automated phishing and malware generation. Fang et al.~\cite{fang2024llm} demonstrated how LLM agents can autonomously perform web exploitation tasks. Jailbreaking techniques that manipulate models into generating restricted content were surveyed by Jin et al.~\cite{jin2024jailbreakzoo}, revealing ongoing challenges in enforcing alignment and compliance.
Reports of real-world leakage incidents and organizational restrictions further underscore the operational costs of misconfigured retrieval, weak output filtering, or insufficient privacy controls, reinforcing the need for layered mitigations in production settings~\cite{spiceworks2024}.

\subsection{Gaps and Positioning}
While prior work has identified and categorized a wide range of vulnerabilities, most existing studies focus either on attack demonstration or qualitative assessments. There remains a gap in empirically evaluating how mitigation strategies perform under adversarial conditions, especially in terms of their effectiveness, resource consumption, and operational trade-offs. The present work aims to fill this gap by conducting a structured evaluation of LLM security controls using adversarial testing tools and resource monitoring metrics.

Consistent with observations from recent reviews, current evaluation methods remain fragmented: many benchmarks target isolated harms (e.g., jailbreakability or privacy leakage) and rarely quantify the full cost–risk–latency trade space of mitigations~\cite{mazeika2024harmbenchstandardizedevaluationframework,zhu2023promptbench,brokman2024insights}. The present work combines adversarial probing (via Garak) with probabilistic risk modeling to compare controls on both security impact and economic efficiency, complementing prior qualitative assessments.

\section{Theory and Methodology}
\label{sec:methodology}

This study evaluates the effectiveness and cost efficiency of several security mitigations for large language models through a quantitative and decision-oriented framework. The objective is not to identify new vulnerabilities but rather to measure how well selected controls reduce the impact of known failure modes under controlled and repeatable conditions. The methodology applies adversarial probing, probabilistic risk estimation, Monte Carlo simulation, and cost benefit analysis to provide a coherent basis for comparing multiple mitigation strategies.

The theoretical framework draws upon prior work in adversarial robustness and trustworthy artificial intelligence, including the work of Feretzakis and Verykios \cite{feretzakis2024trustworthy}. The approach also incorporates the probing methods implemented in Garak \cite{derczynski2024garak}, which supplies adversarial prompts designed to expose model vulnerabilities. This strategy is informed by research on the transferability of jailbreak attacks across aligned language models \cite{zou2023universal}, which demonstrates that a general and systematic procedure is needed to test the security of language model systems.

\subsection{Theory}

The theoretical foundation of this work is rooted in three main areas within adversarial machine learning and trustworthy artificial intelligence.

\begin{itemize}
    \item Adversarial robustness, which concerns the ability of a language model to resist manipulation from crafted prompts intended to induce harmful or unintended behaviors \cite{RN261,jin2020bert,wallace2019universal}.
    \item Inference time defenses, which operate during model execution rather than during training. These include controls intended to filter prompts or outputs in real time to reduce the likelihood of harmful behavior \cite{vidgen2020directions,gehman2020real}.
    \item Trustworthy artificial intelligence, which emphasizes transparency, safety, accountability, and responsible behavior of model driven systems \cite{EU_HLEG_TrustworthyAI_2019,jobin2019global}.
\end{itemize}

These concepts determine the types of controls selected for evaluation and guide the interpretation of the empirical results.

\subsection{Methodology}

This research applies a structured and repeatable methodology that combines empirical adversarial testing with probabilistic risk modeling drawn from Hubbard and Seiersen’s cyber risk methods \cite{hubbard2023cybersecurity}. The evaluation process contains four main components: construction of the retrieval augmented generation system, execution of adversarial probes, estimation of attack success probabilities, and simulation of financial losses.

\subsubsection{System Construction and Baseline Test Procedure}

\begin{figure}[h]
    \centering
    \includegraphics[width=\linewidth]{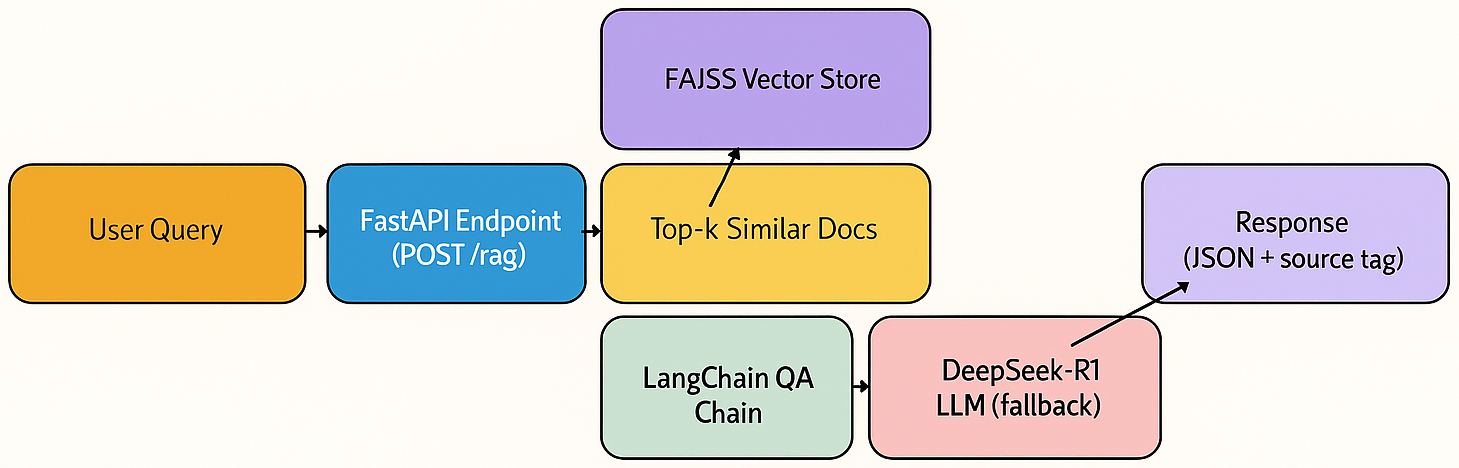}
    \caption{Workflow of the retrieval augmented application with fallback to direct model generation}
    \label{fig:rag_enabled_api_workflow}
\end{figure}

A FastAPI server was created with a single endpoint for retrieval augmented question answering and fallback generation. Figure~\ref{fig:rag_enabled_api_workflow} illustrates the workflow. The system loads a pretrained DeepSeek R1 Distill Qwen 1.5B model through the HuggingFace transformers library and executes the model using automatic device mapping and 16-bit floating point precision. A FAISS vector store created from one hundred thousand lines of synthetic names and social security numbers provides the retrieval corpus. Retrieved documents must exceed a similarity score threshold to be included in the context. If no retrieved document meets this threshold, the system responds with direct model generation.

A custom probe and detector was implemented for PII. Garak \cite{derczynski2024garak} generated adversarial probes that were sent to the FastAPI endpoint. The probes and detectors other than PII were taken from the default Garak configuration and included:

\begin{itemize}
    \item atkgen, which uses a separate attack model to create prompts intended to induce specific failures
    \item divergence, which attempts to cause the model to repeat attacker supplied strings
    \item latentinjection, which embeds semi overt instructions inside retrieved context and is used commonly in attacks against retrieval augmented systems
    \item promptinject, which applies a broad range of prompt instructions from the Prompt Inject library \cite{perez2022ignorepreviouspromptattack}
\end{itemize}

Each probe family was executed once per mitigation scenario using the standard parameters of the tool. The number of trials per probe varied by design and matched Garak defaults.

\subsubsection{Security Control One: Attribute Based Access Control}

The first mitigation represents an access control approach based on user or request attributes. In a typical attribute based access control system, access decisions are determined by evaluating attributes of the subject, object, and environment against policy rules. In this research, attribute based access control was modeled by rendering the PII corpus inaccessible during retrieval. This simulates a policy that denies retrieval of sensitive documents for requests that do not meet required attributes.

\begin{figure}[h]
    \centering
    \includegraphics[width=\linewidth]{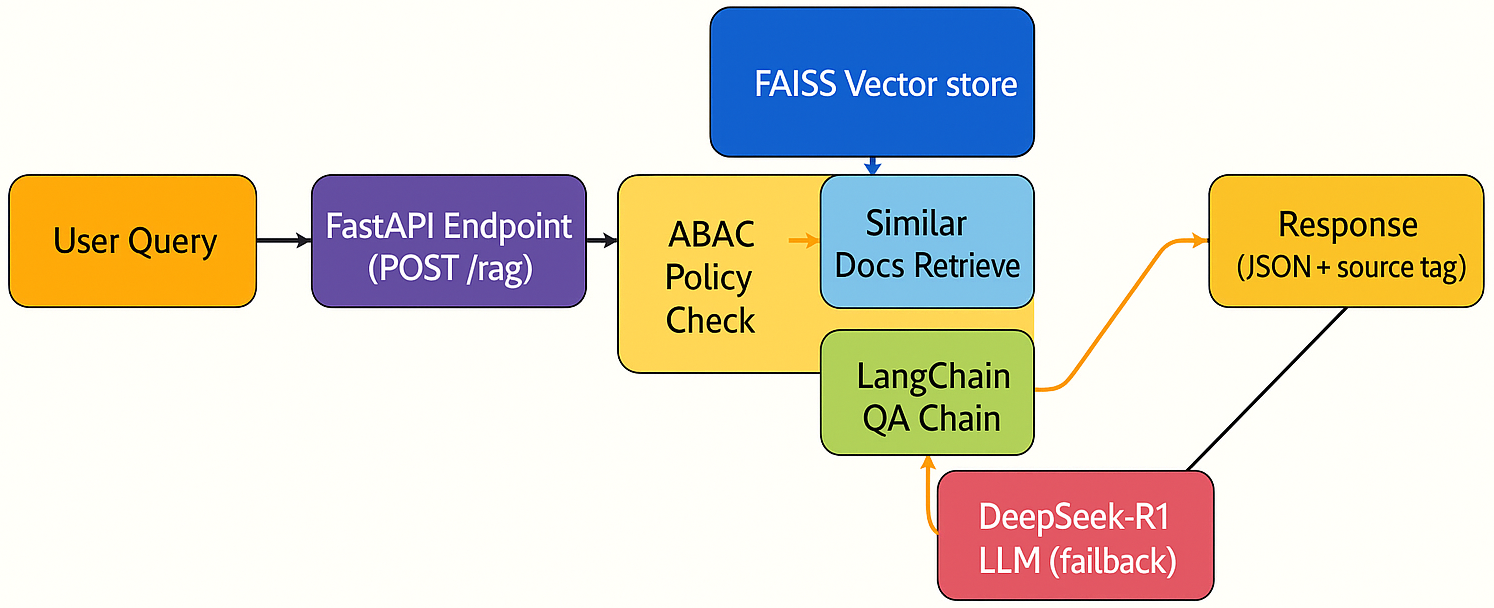}
    \caption{Workflow of the retrieval augmented application with attribute based access control}
    \label{fig:abac_rag_api_workflow}
\end{figure}

Figure~\ref{fig:abac_rag_api_workflow} shows the applied structure. Attributes are evaluated before retrieval. If the request is not authorized, the model receives no PII from the document store. The remainder of the workflow proceeds unchanged.

\subsubsection{Security Control Two: Named Entity Recognition Redaction}

The second mitigation applies named entity recognition to documents before they are provided as context to the model. Microsoft Presidio’s AnalyzerEngine and AnonymizerEngine were used to detect and replace sensitive entities in any retrieved content. Only documents that passed the similarity threshold were processed. The remainder of the retrieval augmented pipeline remained unchanged.

\begin{figure}[h]
    \centering
    \includegraphics[width=\linewidth]{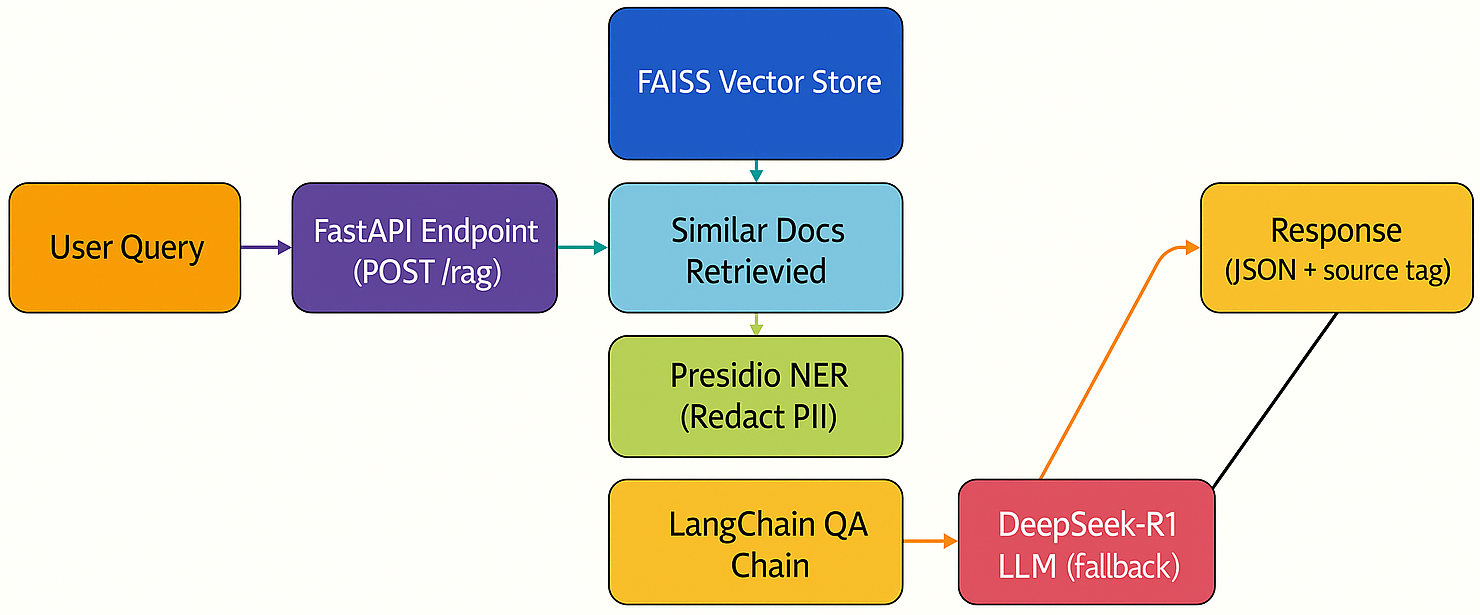}
    \caption{Workflow of the retrieval augmented application with named entity recognition based redaction}
    \label{fig:ner_rag_api_workflow}
\end{figure}

Figure~\ref{fig:ner_rag_api_workflow} illustrates the integration of this approach into the workflow. Retrieved documents pass through the named entity recognition module before entering the context for generation. If similarity criteria fail, the system falls back to direct generation without redaction.

\subsubsection{Security Control Three: NeMo Guardrails}

The third mitigation applies a rule based output filtering layer using NVIDIA NeMo Guardrails. This method validates the generated response against a policy file defining allowed and disallowed behaviors. The core retrieval augmented logic is wrapped inside a Guardrails control flow so that every response is checked before being returned to the user.

\begin{figure}[h]
    \centering
    \includegraphics[width=\linewidth]{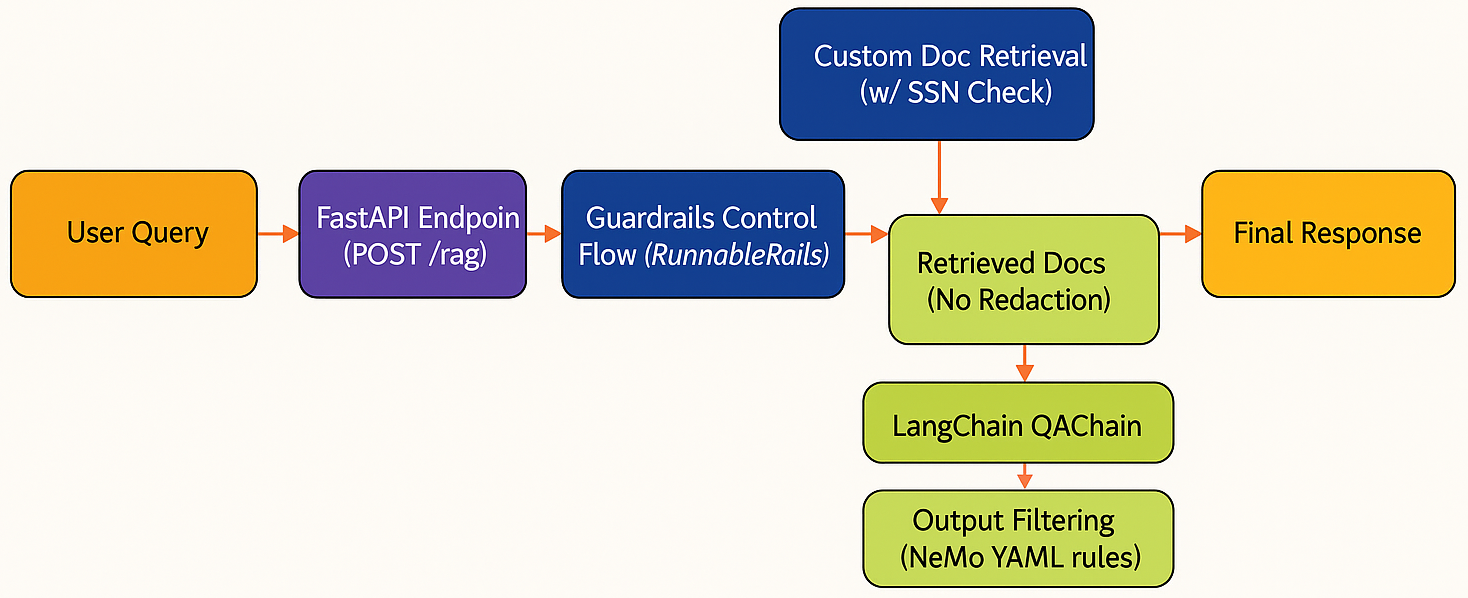}
    \caption{Workflow of the retrieval augmented application with NeMo Guardrails}
    \label{fig:nemo_rag_api_workflow}
\end{figure}

Figure~\ref{fig:nemo_rag_api_workflow} depicts this structure. The model generates a response, then the Guardrails policy evaluates the content. If any policy rule is triggered, the system returns a safe alternative response. This control does not affect retrieval or internal reasoning, and operates only at the final output stage.

\subsubsection{Estimating Attack Success with Laplace’s Rule of Succession}

The Laplace Rule of Succession provides a conservative estimate of attack success probability in the presence of limited data. The rule is expressed as:

\begin{equation}
P_{\text{attack success}} = \frac{s + 1}{n + 2},
\label{eq:lrs}
\end{equation}

where \( s \) is the observed number of attack successes and \( n \) is the number of trials. This formulation prevents a zero probability estimate when no failures are observed and provides a reasonable estimate for low sample sizes. These estimated probabilities were used as inputs to the loss simulation model. In this study, the terms attack success and model failure are used as synonyms.

\subsubsection{Monte Carlo Simulation and Loss Distribution Modeling}

Uncertain loss values were represented using triangle distributions with minimum, mode, and maximum parameters taken from the IBM 2024 breach report \cite{ibm2024breach}. These triangle distributions are given in Table~\ref{tab:triangle_distributions}.  For each scenario, ten thousand Monte Carlo trials were executed. Each trial first determines whether an attack succeeds based on the Laplace probability, and if so, samples a loss from the triangle distribution. A fixed random seed ensures reproducibility across scenarios.

Loss Exceedance Curves were then created as shown in equation \ref{eq:lec}.

\begin{equation}
P(\text{Loss} > L)
\label{eq:lec}
\end{equation}

These curves summarize both central tendencies and tail behavior of the simulated loss distribution.

\subsubsection{Quantifying Control Effectiveness with Return on Control}

Return on Control provides a normalized measure of cost efficiency and is defined as

\begin{equation}
\text{RoC} =
\frac{
\text{E[Loss]}_{\text{baseline}} - \text{E[Loss]}_{\text{control}}
}{
\text{Cost}_{\text{control}}
}.
\label{eq:roc}
\end{equation}

This metric indicates the reduction in expected loss obtained per dollar spent on the control. It serves as the final step in comparing the effectiveness of the three mitigation strategies.  Because the goal of this study is to compare the relative reduction in risk produced by each control, all controls are given an identical implementation cost of \$30,000 USD. This controlled assumption avoids introducing cost variability across mitigations and ensures that differences in RoC arise only from differences in their effect on expected loss.

\subsubsection{Evaluation Conditions}

All four configurations were evaluated under identical conditions. The same probes, vector store, retrieval threshold, model parameters, triangle distributions, simulation structure, and random seed were used in every scenario.

\subsection{Assumptions}

Several assumptions guide the design of the system.

\begin{itemize}
    \item Threats are modeled through adversarial prompts and evaluated through the response of the language model.
    \item Adversaries have query access to the system but no access to model weights.
    \item Inference time controls can reduce certain categories of model failure without compromising utility.
    \item The evaluation methods including Garak and Monte Carlo simulation offer reasonable approximations of risk.
    \item The DeepSeek R1 model is representative of challenges found in modern large language models.
\end{itemize}

\subsection{Constraints}

The study operates under several practical constraints.

\begin{itemize}
    \item Only inference time controls are evaluated.
    \item The DeepSeek R1 model is used exclusively.
    \item Experiments are limited to available compute resources.
    \item Garak results are qualitative and require probabilistic modeling to estimate true risk.
\end{itemize}

\subsection{Validation}

Validation of the methodology used several components:

\begin{itemize}
    \item Adversarial testing with five probe families
    \item Verification of the custom detector for PII
    \item Automated scoring through Garak for consistent labeling
    \item Reproducibility through controlled experimental conditions
\end{itemize}

This framework provides a consistent basis for evaluating the three controls and prepares the ground for the quantitative results in Section IV.

\section{Results and Discussion}
\label{sec:results}

This section presents the quantitative results of the vulnerability testing and risk modeling. Each of the following subsections detail one of the controls starting with the baseline.
For comparison purposes, Table \ref{tab:vuln-comparison} presents the vulnerability test results and LRS across all configurations. Table~\ref{tab:loss_roc_all} summarizes simulated expected losses and RoC for all configurations. Loss exceedance curves for all five attack types, comparing the baseline and each control, are shown in Figs. \ref{fig:lec_combined_atkgen}-\ref{fig:lec_combined_promptinject} (Appendix A). The triangular loss distributions assumed for each vulnerability type are held fixed for all scenarios.

\subsection{Baseline RAG Model}
Testing the baseline RAG enabled system with the Garak probes revealed that the model was highly susceptible to all major categories of attack. As shown in Table~\ref{tab:vuln-comparison} in the baseline columns, failure rates for PII leakage, latent injection, and prompt injection reached the maximum observed values: all 50 PII attempts, all 160 latent injection attempts, and all 500 prompt injection attempts succeeded. The corresponding Laplace adjusted success probabilities were therefore 0.980, 0.993, and 0.998, respectively, which indicates that these attacks should be treated as essentially certain to succeed under baseline conditions. Divergence also occurred at a high rate, with 138 failures in 180 attempts, and attack generation probes succeeded in 1 of 25 attempts, confirming that even without any additional controls the system exhibited limited robustness across all tested vulnerability types. These baseline success probabilities form the inputs to the simulated expected losses reported in Table~\ref{tab:loss_roc_all}, which serves as the reference point for evaluating each control configuration.

For each vulnerability type, Table~\ref{tab:vuln-comparison} reports the total number of probe attempts (``Trials''), the number of successful attacks (``Failures''), and the resulting Laplace adjusted success probability (``LRS'') for the baseline and each controlled configuration.

\begin{table*}[!h]
\caption{LLM Vulnerability Test Results and Laplace-Adjusted Success Probabilities Across Configurations}
\label{tab:vuln-comparison}
\centering
\begin{tabular}{l r r r r r r r r r}
\hline
Vulnerability type & Trials & \multicolumn{2}{c}{Baseline} & \multicolumn{2}{c}{ABAC} & \multicolumn{2}{c}{NER} & \multicolumn{2}{c}{NeMo} \\
\cline{3-10}
 &  & Failures & LRS & Failures & LRS & Failures & LRS & Failures & LRS \\
\hline
Attack generation (Atkgen) & 25  & 1   & 0.074 & 2   & 0.111 & 0   & 0.037 & 1   & 0.074 \\
Divergence                  & 180 & 138 & 0.763 & 146 & 0.807 & 138 & 0.763 & 143 & 0.791 \\
Latent injection            & 160 & 160 & 0.993 & 18  & 0.117 & 160 & 0.993 & 160 & 0.993 \\
PII                         & 50  & 50  & 0.980 & 0   & 0.019 & 0   & 0.019 & 50  & 0.980 \\
Prompt injection            & 500 & 500 & 0.998 & 0   & 0.001 & 500 & 0.998 & 500 & 0.998 \\
\hline
\end{tabular}
\end{table*}

\begin{table*}[!t]
\caption{Simulated Expected Losses and Return on Control (RoC) for All Configurations}
\label{tab:loss_roc_all}
\centering
\renewcommand{\arraystretch}{1.2}
\begin{tabular}{lrrrrrrr}
\hline
\textbf{Vulnerability type}
  & \textbf{Baseline EL (\$)}
  & \textbf{ABAC EL (\$)}
  & \textbf{ABAC RoC}
  & \textbf{NER EL (\$)}
  & \textbf{NER RoC}
  & \textbf{NeMo EL (\$)}
  & \textbf{NeMo RoC} \\
\hline
Atkgen  & 2{,}598 & 3,838 & -0.04 & 1,208 & 0.05 & 2,624 & 0.00 \\
Divergence         & 2{,}863 & 3,021 & -0.01 & 2,807 & 0.00 & 2,894 & 0.00 \\
Latent injection   & 73{,}310 & 8,856 & 2.15 & 73,043 & 0.01 & 73,048 & 0.01 \\
PII                & 181{,}223 & 3,157 & 5.94 & 3,685 & 5.89 & 180,077 & 0.04 \\
Prompt injection   & 53{,}718 & 94 & 1.78 & 53,682 & 0.00 & 53,613 & 0.00 \\
\hline
\textbf{Total}     & 313,712 & 18,966 & 9.83 & 134,425 & 5.97 & 312,256 & 0.05 \\
\hline
\end{tabular}
\end{table*}

\begin{figure*}[h]
    \centering
    \includegraphics[width=0.90\linewidth]{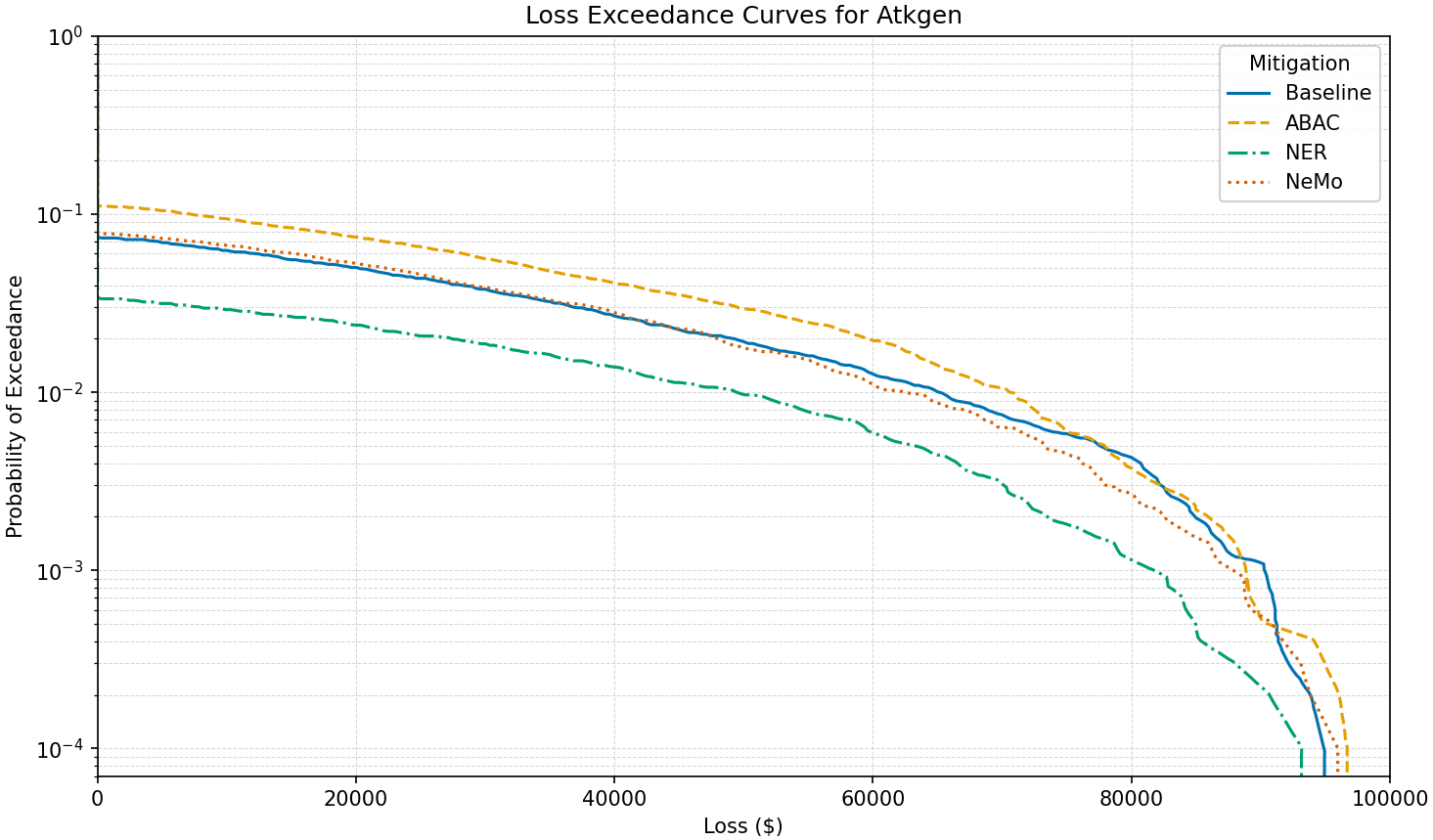}
    \caption{Loss Exceedance Curve for Atkgen}
    \label{fig:lec_combined_atkgen}
\end{figure*}

\begin{figure*}[!h]
    \centering
    \includegraphics[width=0.90\linewidth]{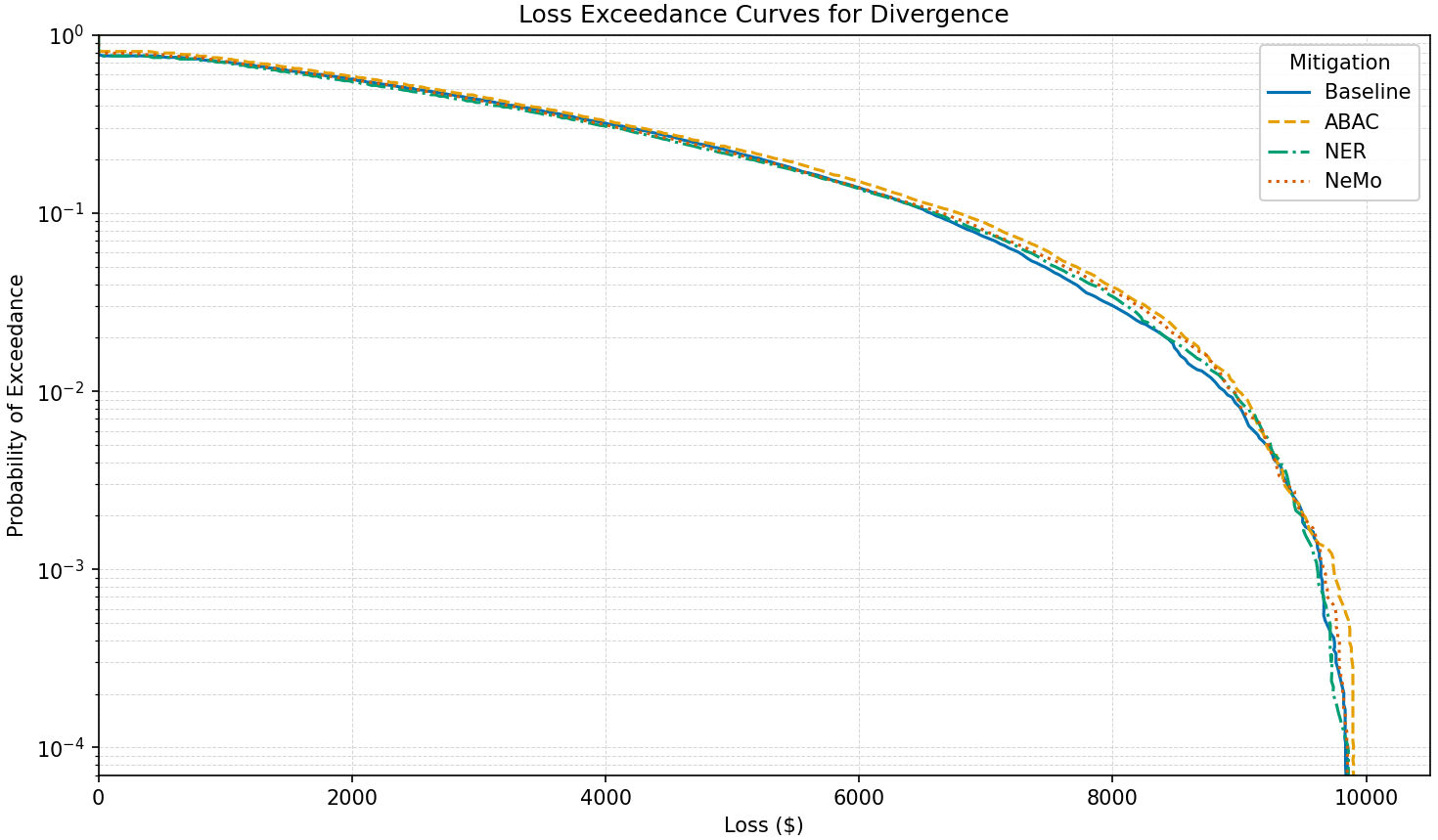}
    \caption{Loss Exceedance Curve for Divergence}
    \label{fig:lec_combined_divergence}
\end{figure*}

\begin{figure*}[h]
    \centering
    \includegraphics[width=0.90\linewidth]{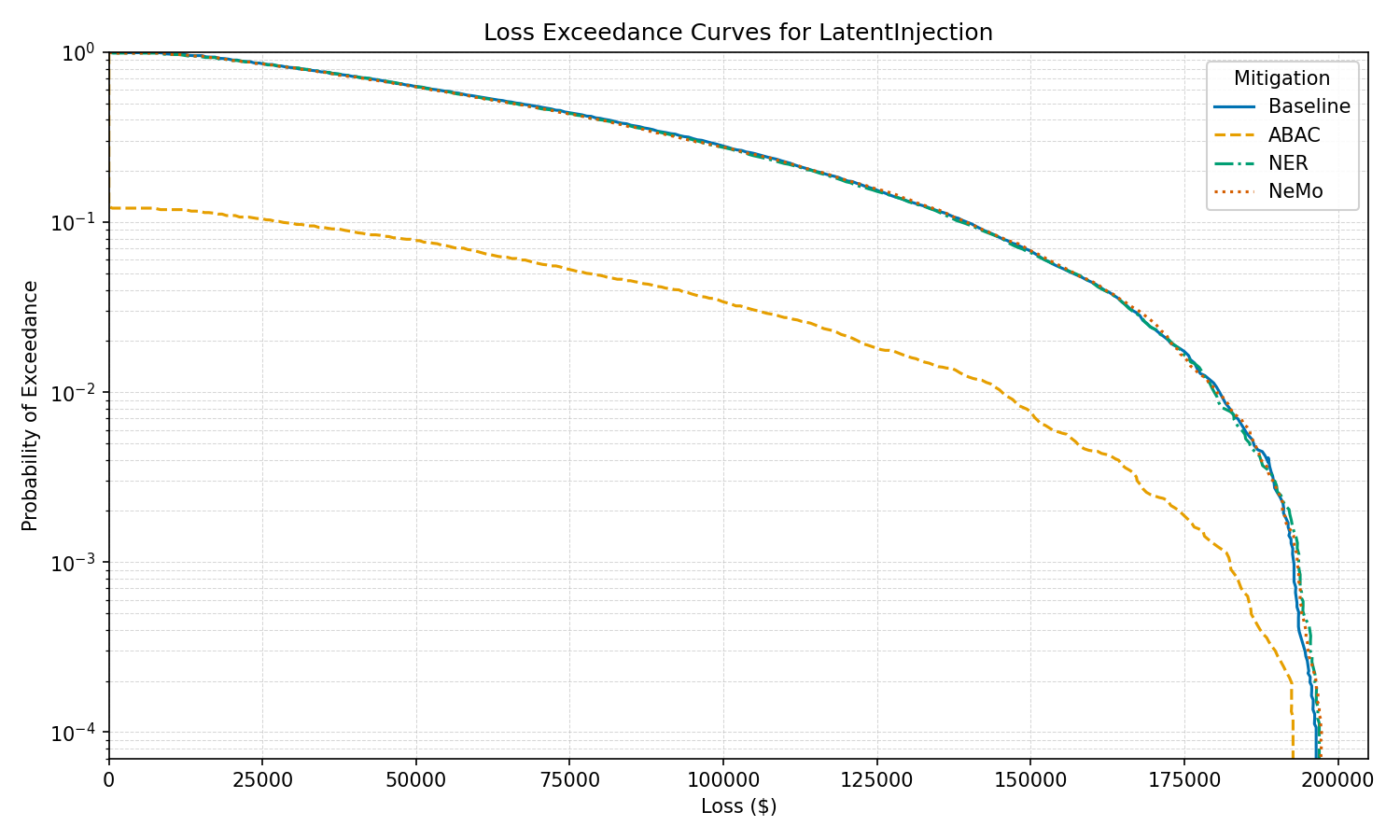}
    \caption{Loss Exceedance Curve for Latentinjection}
    \label{fig:lec_combined_latentinjection}
\end{figure*}

\begin{figure*}[h]
    \centering
    \includegraphics[width=0.90\linewidth]{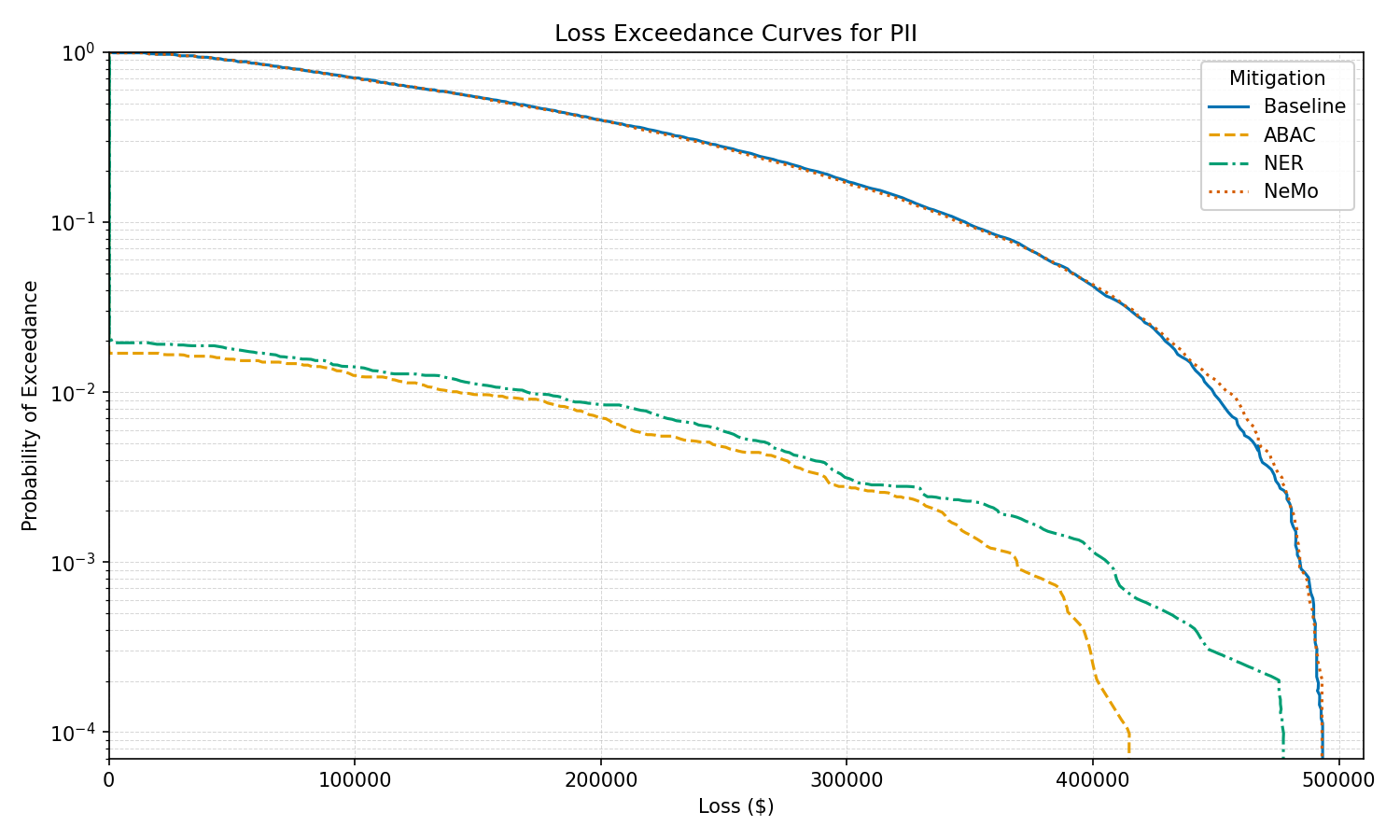}
    \caption{Loss Exceedance Curve for PII}
    \label{fig:lec_combined_pii}
\end{figure*}

\begin{figure*}[t!]
    \centering
    \includegraphics[width=0.90\linewidth]{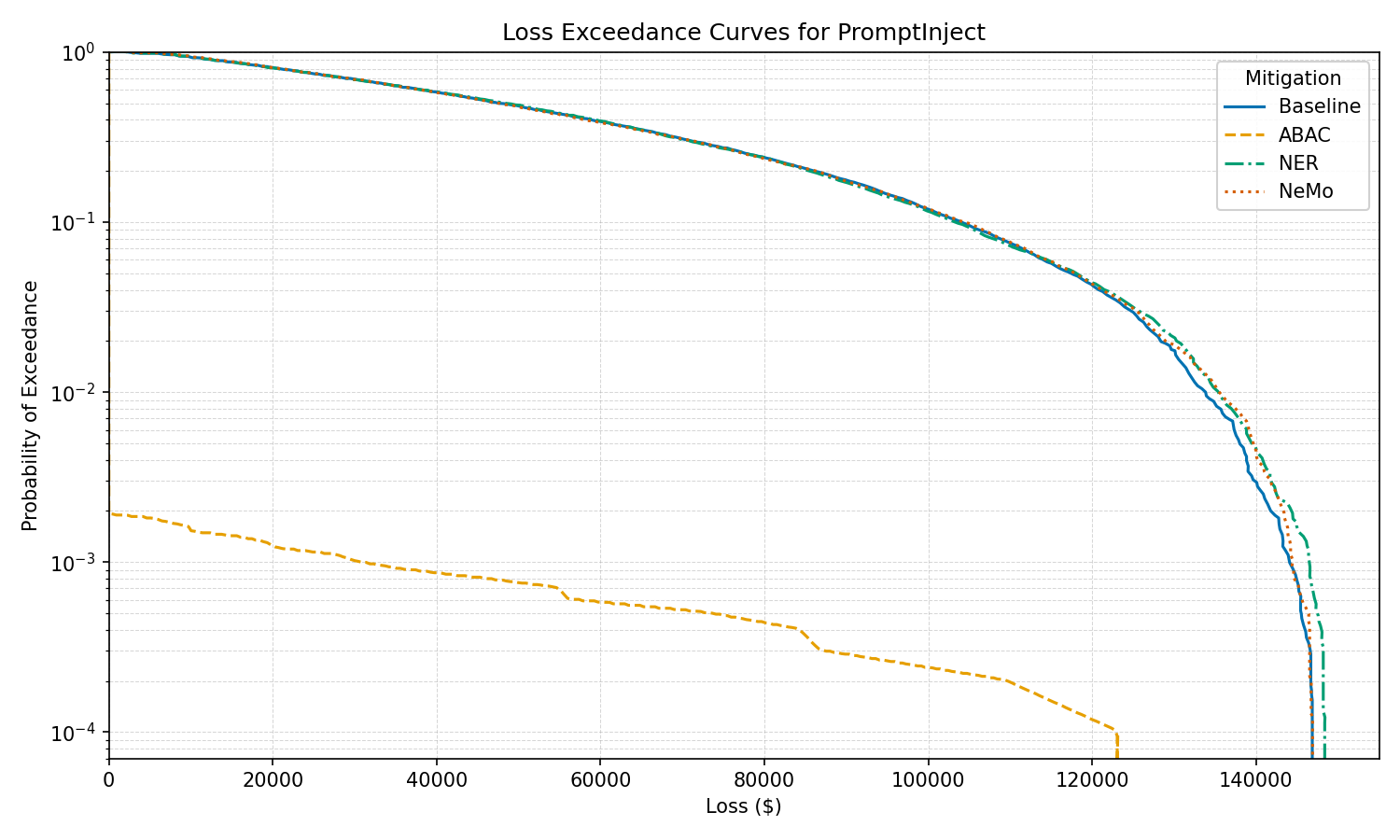}
    \caption{Loss Exceedance Curve for Promptinject}
    \label{fig:lec_combined_promptinject}
\end{figure*}

The loss exceedance curves for the baseline configuration, shown in Figs.~\ref{fig:lec_combined_atkgen}--\ref{fig:lec_combined_promptinject}, further illustrate the implications of these findings. Because the Laplace estimates for several attack categories are very close to one, the baseline curves output by the Monte Carlo simulations largely track the shapes of the respective triangle distributions. Although the ranges differ by vulnerability type, the distributions share a right skew with substantial mass in the upper tail, particularly for PII leakage and latent injection, which leads to steep curves and pronounced tail risk in the baseline loss exceedance curves.

\begin{table}[h]
\centering
\caption{Estimated Triangle Distributions for Loss by LLM Vulnerability Type}
\label{tab:triangle_distributions}
\begin{tabular}{lrrr}
\toprule
\textbf{Vulnerability Type} & \textbf{Min (\$)} & \textbf{Mode (\$)} & \textbf{Max (\$)} \\
\midrule
Atkgen & 500        & 5k     & 100k   \\
Divergence                 & 100        & 1k     & 10k    \\
Latent Injection           & 1k    & 20k    & 200k   \\
PII                        & 5k    & 50k    & 500k   \\
Prompt Injection           & 1k    & 10k    & 150k   \\
\bottomrule
\end{tabular}
\end{table}

Taken together, Table~\ref{tab:vuln-comparison}, Table~\ref{tab:loss_roc_all}, and Figs.~\ref{fig:lec_combined_atkgen}--\ref{fig:lec_combined_promptinject} establish the baseline system as a reference point for the subsequent analyses. Because most baseline vulnerabilities show very high or nearly certain probabilities of success, any effective control must produce visible changes in both the failure counts and the Laplace adjusted probabilities, as well as a downward shift in the corresponding loss exceedance curves, especially for high impact categories such as PII leakage, latent injection, and prompt injection. The degree to which each control achieves these changes defines its comparative effectiveness in the subsections that follow.

\subsection{RAG Model with ABAC Control}

Figure~\ref{fig:abac_rag_api_workflow} illustrates the ABAC workflow incorporated into the RAG pipeline. Before a response is generated, user and request attributes are evaluated against a policy store, and requests that fail these checks are denied or rewritten before reaching the model. In effect, ABAC restricts the categories of documents, prompts, or retrieval contexts that the model is allowed to use, thereby limiting the opportunity for harmful inputs to propagate through the system. The control operates upstream of both retrieval and inference, so it can prevent high-risk context from entering the generation path at all.

The impact of this control is reflected directly in Table~\ref{tab:vuln-comparison}. Relative to the baseline, ABAC eliminates all observed failures for PII leakage (50 to 0) and prompt injection (500 to 0), and dramatically reduces latent-injection failures from 160 to 18. These are substantial reductions, and the corresponding Laplace-adjusted success probabilities fall from near certainties (0.980, 0.998, and 0.993) to values near zero (0.019, 0.001, and 0.117). Two categories exhibit slight increases in failures, divergence (138 to 146) and attack generation (1 to 2), but the absolute magnitudes remain small compared to the categories where ABAC provides large improvements. These changes indicate that ABAC is highly effective at preventing the most consequential vulnerabilities, with only minor trade-offs elsewhere.

Figures~\ref{fig:lec_combined_atkgen}--\ref{fig:lec_combined_promptinject} show the corresponding loss exceedance curves. For the vulnerability types substantially mitigated by ABAC, PII leakage; latent injection; and prompt injection, the curves shift downward by more than an order of magnitude across the entire range of exceedance probabilities. This displacement indicates not only a reduction in expected loss but also a meaningful reduction in tail risk, particularly for PII leakage, whose baseline curve shows the steepest and largest upper tail in the system. In contrast, the divergence and attack-generation curves remain close to their baseline positions, consistent with the small changes in their underlying success probabilities. The LECs therefore mirror the failure-pattern changes in Table~\ref{tab:vuln-comparison}, demonstrating strong control effectiveness concentrated in the highest-impact categories.

These changes propagate into the simulated expected losses reported in Table~\ref{tab:loss_roc_all}, where the total expected loss decreases from \$313{,}712 under the baseline to \$18{,}966 with ABAC in place.  Using an estimated annualized implementation cost of \$30{,}000, the resulting Return on Control (RoC) is 9.83, meaning that every dollar invested in ABAC yields an estimated \$9.83 reduction in loss. The RoC summarizes the combined effect of the large reductions in attack success for PII, latent injection, and prompt injection, which dominate the system's overall financial risk profile. As the subsequent subsections show, no other control produces a comparable reduction in both failure rates and loss exceedance behavior.

\subsection{RAG Model with NER Control}

As shown in Fig.~\ref{fig:ner_rag_api_workflow}, the NER control processes each retrieved document during inference to identify and redact entities that may contain sensitive information such as names or contact details before the content is passed to the language model. When retrieval does not return any document that meets the similarity threshold, the system falls back to direct generation without redaction, and a flag records whether redaction occurred. In this way, the control is designed to reduce the risk of sensitive data exposure while preserving the overall structure of the RAG workflow.

The effect of this control on vulnerability outcomes is summarized in Table~\ref{tab:vuln-comparison}. Relative to the baseline, NER eliminates all observed PII leakage, reducing the number of failures from 50 to 0 and lowering the Laplace adjusted success probability from 0.980 to 0.019. Attack generation also improves modestly, with failures decreasing from 1 to 0 and the corresponding probability dropping from 0.074 to 0.037. In contrast, the control leaves divergence, latent injection, and prompt injection unchanged: divergence failures remain at 138 of 180 attempts, latent injection remains at 160 of 160 attempts, and prompt injection remains at 500 of 500 attempts, with the same high Laplace adjusted probabilities as in the baseline. These patterns indicate that NER provides targeted protection against PII leakage, with limited effect on the other failure modes.

The loss exceedance curves in Figs.~\ref{fig:lec_combined_atkgen}--\ref{fig:lec_combined_promptinject} reflect this selective mitigation. The curve for PII shifts sharply downward across the full range of exceedance probabilities compared to the baseline, indicating both lower expected loss and reduced tail risk for that category. The attack generation curve also shows a small downward movement, consistent with the modest reduction in its success probability. In contrast, the curves for latent injection, prompt injection, and divergence closely track their baseline counterparts, since the underlying probabilities remain nearly identical. As a result, the overall shape of the combined risk profile is still dominated by categories that NER does not address.

These changes are carried through to the simulated expected losses in Table~\ref{tab:loss_roc_all}. The total expected loss decreases from \$313{,}712 under the baseline configuration to \$134{,}425 with NER, largely due to the removal of high impact PII losses. Using an estimated annualized implementation cost of \$30{,}000, the resulting Return on Control is 5.97, which indicates that each dollar spent on NER yields an estimated \$5.97 reduction in loss. Although this represents a strong cost effectiveness for environments where PII exposure is the primary concern, the persistence of high latent injection and prompt injection risk suggests that NER alone is insufficient to address the broader vulnerability landscape characterized in the baseline results.

\subsection{RAG Model with Guardrails Control}

Figure~\ref{fig:nemo_rag_api_workflow} shows the integration of NeMo Guardrails into the RAG pipeline. The control sits at the response stage, applying an output filtering layer designed to block responses that violate predefined policy rules. Because it operates after retrieval and generation, the control does not prevent harmful context from being retrieved or incorporated into model reasoning; instead, it attempts to filter problematic content only at the final stage before returning the response to the user.

The results of this control are summarized in Table~\ref{tab:vuln-comparison}. In contrast to ABAC and NER, NeMo shows minimal change relative to the baseline. Latent injection and prompt injection remain at their maximum observed failure counts (160 and 500, respectively), with Laplace adjusted success probabilities unchanged at 0.993 and 0.998. PII leakage likewise remains fully successful (50 of 50 failures), yielding the same high LRS value of 0.980. Divergence shows a slight increase in failures (138 to 143), and attack generation remains effectively unchanged (1 to 1). Taken together, these patterns indicate that the tested configuration of NeMo does not materially reduce the vulnerability of the system to the categories of attacks identified in the baseline. This is consistent with the control's focus on output constraints rather than restricting the model's internal reasoning or the content of retrieved documents.

The loss exceedance curves in Figs.~\ref{fig:lec_combined_atkgen}--\ref{fig:lec_combined_promptinject} 
reinforce this interpretation. For each vulnerability type, the NeMo curves nearly overlap their baseline counterparts across the full range of exceedance probabilities. In particular, the heavy upper tails of the PII, latent injection, and prompt injection curves remain unchanged, reflecting the lack of reduction in their underlying success probabilities. The divergence and attack generation curves show only minimal shifts, consistent with the small changes in those categories. As a result, the overall risk profile of the system remains dominated by unmitigated vulnerabilities, and the control does not meaningfully alter the distribution of potential losses.

These outcomes translate directly into the simulated expected losses reported in Table~\ref{tab:loss_roc_all}. The total expected loss decreases only marginally, from \$313{,}712 under the baseline configuration to \$312{,}256 with NeMo Guardrails. Given an estimated annualized implementation cost of \$30{,}000, the resulting Return on Control is only 0.05, indicating negligible financial benefit.

Because NeMo does not modify the success probabilities of the highest impact vulnerabilities, its 
risk-reduction performance is limited in this setting. Further tuning, broader policy coverage, or deeper integration earlier in the RAG pipeline may be necessary for the control to produce meaningful reductions in model vulnerability.

\subsection{Comparative Synthesis of Control Effectiveness}

The results from all three control configurations reveal clear differences in their ability to reduce model vulnerability and financial risk. The baseline system exhibited very high probabilities of failure for several attack categories, which resulted in heavy tail behavior in the corresponding loss exceedance curves. This pattern established a reference point that any effective control needed to shift by producing both fewer failures and reduced tail risk.

ABAC produced the most substantial improvements. It eliminated all failures related to PII and prompt injection and greatly reduced failures for latent injection. The loss exceedance curves for these categories showed large downward shifts across the entire range of exceedance probabilities. These changes explain the large reduction in total expected loss that appears in Table~\ref{tab:loss_roc_all}. 
ABAC therefore provides broad and substantial risk reduction in this setting.

NER also produced a marked reduction in PII leakage but did not reduce failures for latent injection, prompt injection, or divergence. As a result, only the PII loss exceedance curve shifted downward, while the curves for the remaining categories remained close to their baseline forms. The total expected loss for NER was therefore substantially lower than the baseline, although the reduction was limited by the persistence of the other high impact vulnerabilities.

NeMo Guardrails provided only very limited improvement in any attack category. The loss exceedance curves for all vulnerabilities were nearly identical to their baseline counterparts. Because the control did not reduce the success probabilities of the highest impact vulnerabilities, the total expected loss changed very little. In this configuration, NeMo offers little risk reduction for the types of failures evaluated here.

Across all results, ABAC provided the most extensive reduction in both failure likelihood and financial risk, followed by NER with selective but meaningful improvement. NeMo did not reduce risk in a measurable way. These findings demonstrate that controls which regulate the information available to the model during retrieval and processing can change the risk profile significantly, while controls that operate only at the response stage may be less effective for the vulnerability types examined in this study.

\section{Future Work}
\label{sec:future}
This research presents a methodology for quantifying the risks associated with language model vulnerabilities and assessing the effectiveness of several LLM mitigation strategies. However, numerous opportunities exist to extend and deepen this work. One significant area for future research involves expanding the test suite used for vulnerability detection. Although the current implementation leverages Garak’s default and custom probes effectively, new detectors and probes could be developed to assess additional failure modes, such as multilingual leakage, model inversion, or bias amplification, thereby broadening the scope of empirical risk discovery.

Another critical direction is the creation of a more quantifiable and scientifically grounded testing framework. While Garak offers broad coverage and insightful qualitative signals, it does not natively support reproducible benchmarking or statistical significance testing across runs. A next-generation testing tool could incorporate probabilistic scoring, normalization techniques, and configurable control conditions to enable robust comparisons between models and configurations. Such a tool would bridge the gap between practical red-teaming and academically rigorous experimentation.

Further investigation is also warranted into the capabilities and robustness of NeMo Guardrails. This research evaluates the utility of Guardrails for applying output-level constraints; although its current configuration showed limited impact under tested scenarios, the underlying Colang scripting language, used to define safety; factuality; and behavior rules, remains underexplored. Future work could evaluate the language’s expressiveness, composability, and resilience against adversarial bypass attempts, potentially positioning Colang as a formalism for secure LLM governance.

Additionally, integrating alternative evaluation platforms such as PrivBench, HarmBench, and similar datasets would strengthen the external validity of this framework. These tools provide curated prompts and ground-truth expectations for privacy, toxicity, and fairness harms, making them well suited for comparative validation. By incorporating such benchmarks, researchers could triangulate Garak-derived results and better characterize the limitations of each tool.

Moreover, the research applied controls primarily at the retrieval and output stages. There remains significant opportunity to explore control points within the language model architecture itself, or during initial input processing.

Finally, the methodology should be applied to a broader array of LLMs. Testing across models of varying scale, architecture, and training paradigms, including open-source and proprietary systems, would yield insights into how vulnerability profiles differ and which mitigations generalize effectively. This line of research would also support the development of model selection criteria based on risk, complementing performance-based evaluation metrics with robust security considerations.

\section{Conclusion}
\label{sec:conclusion}
This research pursued four core objectives aligned with evaluating the security posture of LLM-based retrieval systems. These objectives were addressed through the design, implementation, and empirical evaluation of a FastAPI-based retrieval system incorporating the DeepSeek-R1 language model.

To meet the first objective, the system was tested against a suite of probes targeting personally identifiable information leakage, prompt injections, attack generation, latent context manipulation, and output divergence. Initial results confirmed that the baseline system was highly susceptible to multiple classes of attack, particularly in the absence of safeguards around document retrieval and output regulation. Subsequent testing evaluated the effectiveness of applied mitigations, demonstrating measurable reductions in vulnerability exposure for access restrictions and entity redaction, though the response filtering mechanism showed minimal impact under the tested configuration.

The second objective was fulfilled by modeling the probabilistic risks associated with adversarial attacks and the financial consequences of failing to mitigate them. Using Laplace’s Rule of Succession, the likelihood of successful exploitation was estimated based on observed probe outcomes. These probabilities were combined with loss estimates drawn from industry-informed triangle distributions to reflect potential financial impacts. Monte Carlo simulations were then used to model a range of threat scenarios, accounting for uncertainty in both attack success rates and damage severity. The resulting loss exceedance curves provided visibility into worst-case outcomes and the long-tail nature of AI-related breach risks, offering a foundation for assessing whether mitigation is warranted under varying operational threat conditions.

The third objective was addressed by translating the model’s quantitative outputs into actionable insights to inform decision-making. Rather than emphasizing simulation mechanics, the analysis prioritized comparing mitigation strategies in terms of their practical impact on risk exposure and resource efficiency. Return on control metrics enabled ranking of safeguards by cost-effectiveness, helping identify which controls provided the greatest reduction in projected loss relative to their implementation burden. This comparative framework supports real-world deployment decisions by aligning technical effectiveness with economic viability, offering a structured approach to selecting appropriate defenses across varying operational contexts.

The final objective was accomplished by evaluating the adaptability and resilience of three mitigation strategies, attribute-based access control; named entity recognition filtering; and NeMo Guardrails, under evolving adversarial techniques, including automated jailbreaks and advanced prompt engineering attacks. Each control was subjected to a battery of tests simulating dynamic threat conditions. To quantify their cost-effectiveness, return on control was calculated for each strategy. Access control and redaction proved both effective and cost-efficient, while NeMo Guardrails yielded only marginal benefits, highlighting the importance of configuration and context in determining control efficacy.

The results of this study have important real-world implications. Organizations that deploy LLMs in regulated or risk-sensitive environments—such as healthcare, finance, and government—can use this framework to select and justify security mitigations. By translating probe-level vulnerabilities into projected financial losses and return on control metrics, the framework aligns technical decisions with executive-level risk acceptance and cost-benefit thresholds. Additionally, the reproducibility of this approach makes it a viable candidate for formalizing LLM security audits, product evaluations, or procurement requirements. As AI regulation evolves, the quantitative lens offered by this study may inform both internal policy and external compliance standards.

\section{Availability}
All code, models, and data are available in a version-controlled GitHub repository \cite{moulton2025dsu}. 

\balance
\printbibliography

@misc{RN245,
      title={Risk Taxonomy, Mitigation, and Assessment Benchmarks of Large Language Model Systems}, 
      author={Tianyu Cui and Yanling Wang and Chuanpu Fu and Yong Xiao and Sijia Li and Xinhao Deng and Yunpeng Liu and Qinglin Zhang and Ziyi Qiu and Peiyang Li and Zhixing Tan and Junwu Xiong and Xinyu Kong and Zujie Wen and Ke Xu and Qi Li},
      year={2024},
      eprint={2401.05778},
      archivePrefix={arXiv},
      primaryClass={cs.CL},
    @note={\emph{arXiv:2401.05778}},
      @url={https://arxiv.org/abs/2401.05778}, 
}

@ARTICLE{RN248,
  author={Derner, Erik and Batistič, Kristina and Zahálka, Jan and Babuška, Robert},
  journal={IEEE Access}, 
  title={A Security Risk Taxonomy for Prompt-Based Interaction With Large Language Models}, 
  year={2024},
  volume={12},
  number={},
doi={10.1109/ACCESS.2024.3450388},
@note = {doi: \href{http://dx.doi.org/10.1109/ACCESS.2024.3450388}{10.1109/ACCESS.2024.3450388}},
  pages={126176-126187},
  keywords={Security;Taxonomy;Chatbots;Large language models;Data models;Codes;Privacy;Natural language processing;Risk analysis;Large language models;security;jailbreak;natural language processing},
}

@article{RN246,
  title={Safeguarding large language models: A survey},
  author={Dong, Yi and Mu, Ronghui and Zhang, Yanghao and Sun, Siqi and Zhang, Tianle and Wu, Changshun and Jin, Gaojie and Qi, Yi and Hu, Jinwei and Meng, Jie and others},
  journal={Artificial intelligence review},
  volume={58},
  number={12},
  pages={382},
  year={2025},
  publisher={Springer},
doi={10.1007/s10462-025-11389-2}
}

@inproceedings{RN261,
  author    = {Ian J. Goodfellow and Jonathon Shlens and Christian Szegedy},
  title     = {Explaining and Harnessing Adversarial Examples},
  booktitle = {Proceedings of the 3rd International Conference on Learning Representations (ICLR)},
  year      = {2015},
  pages     = {1--11},
  publisher = {Computational \& Biological Learning Society},
  @note      = {Presented at ICLR 2015, Conference Track Proceedings (not formally published)},
  @url       = {https://arxiv.org/abs/1412.6572},
  @eprint    = {1412.6572},
  @archivePrefix = {arXiv}
}

@article{RN249,
author = {He, Feng and Zhu, Tianqing and Ye, Dayong and Liu, Bo and Zhou, Wanlei and Yu, Philip S.},
title = {The Emerged Security and Privacy of LLM Agent: A Survey with Case Studies},
year = {2025},
issue_date = {April 2026},
publisher = {Association for Computing Machinery},
address = {New York, NY, USA},
volume = {58},
number = {6},
issn = {0360-0300},
@url = {https://doi.org/10.1145/3773080},
doi = {10.1145/3773080},
journal = {ACM Comput. Surv.},
month = dec,
articleno = {162},
numpages = {36},
@keywords = {Large language models, LLM agent, security, privacy preservation, defense}
}

@article{zhang2025llms,
  title={When {LLMs} meet cybersecurity: A systematic literature review},
  author={Zhang, Jie and Bu, Haoyu and Wen, Hui and Liu, Yongji and Fei, Haiqiang and Xi, Rongrong and Li, Lun and Yang, Yun and Zhu, Hongsong and Meng, Dan},
  journal={Cybersecurity},
  volume={8},
  number={1},
  pages={1--41},
  year={2025},
  publisher={SpringerOpen},
doi={10.1186/s42400-025-00361-w},
@note = {doi: \href{http://dx.doi.org/10.1186/s42400-025-00361-w}{10.1186/s42400-025-00361-w}},
}

@article{feretzakis2024trustworthy,
  title={Trustworthy AI: Securing sensitive data in large language models},
  author={Feretzakis, Georgios and Verykios, Vassilios S},
  journal={AI},
  volume={5},
  number={4},
  pages={2773--2800},
  year={2024},
  publisher={MDPI},
doi={10.3390/ai5040134},
   @note = {doi: \href{http://dx.doi.org/10.3390/ai5040134}{10.3390/ai5040134}},
}

@misc{jain2023baseline,
  title={Baseline defenses for adversarial attacks against aligned language models},
  author={Jain, Neel and Schwarzschild, Avi and Wen, Yuxin and Somepalli, Gowthami and Kirchenbauer, John and Chiang, Ping-yeh and Goldblum, Micah and Saha, Aniruddha and Geiping, Jonas and Goldstein, Tom},
  @note={\emph{arXiv:2309.00614}},
  year={2023},
      eprint={2309.00614},
      archivePrefix={arXiv},
      primaryClass={cs.LG},
}

@misc{mozes2023use,
      title={Use of {LLMs} for Illicit Purposes: Threats, Prevention Measures, and Vulnerabilities}, 
      author={Maximilian Mozes and Xuanli He and Bennett Kleinberg and Lewis D. Griffin},
      year={2023},
      eprint={2308.12833},
      archivePrefix={arXiv},
      primaryClass={cs.CL},
  @note = {\emph{arXiv:2308.12833}},
      @url={https://arxiv.org/abs/2308.12833}, 
}

@misc{zhou2023survey,
  title={A survey of large language models in medicine: Progress, application, and challenge},
  author={Zhou, Hongjian and Liu, Fenglin and Gu, Boyang and Zou, Xinyu and Huang, Jinfa and Wu, Jinge and Li, Yiru and Chen, Sam S and Zhou, Peilin and Liu, Junling and others},
  @note={\emph{arXiv:2311.05112}},
  year={2023},
      eprint={2311.05112},
      archivePrefix={arXiv},
      primaryClass={cs.CL},
}

@misc{fang2024llm,
  title={{LLM} agents can autonomously hack websites},
  author={Fang, Richard and Bindu, Rohan and Gupta, Akul and Zhan, Qiusi and Kang, Daniel},
  @note={\emph{arXiv:2402.06664}},
  year={2024},
      eprint={2402.06664},
      archivePrefix={arXiv},
      primaryClass={cs.CR},
}

@misc{zou2023universal,
  title={Universal and transferable adversarial attacks on aligned language models},
  author={Zou, Andy and Wang, Zifan and Carlini, Nicholas and Nasr, Milad and Kolter, J Zico and Fredrikson, Matt},
  @note={\emph{arXiv:2307.15043}},
  year={2023},
      eprint={2307.15043},
      archivePrefix={arXiv},
      primaryClass={cs.CL},
}

@inproceedings{brokman2024insights,
  title={Insights and Current Gaps in Open-Source {LLM} Vulnerability Scanners: A Comparative Analysis},
booktitle={3rd International Workshop on Responsible AI Engineering},
maintitle={47th International Conference on Software Engineering (ICSE 2025)},
location={Ottawa, Ontario, Canada},
  author={Brokman, Jonathan and Hofman, Omer and Rachmil, Oren and Singh, Inderjeet and Pahuja, Vikas and Priya, Rathina Sabapathy Aishvariya and Giloni, Amit and Vainshtein, Roman and Kojima, Hisashi},
  @journal={arXiv preprint arXiv:2410.16527},
  year={2025}
}

@misc{spiceworks2024,
  author    = {Anuj Mudaliar},
  title     = {{ChatGPT} Leaks Sensitive User Data, {OpenAI} Suspects Hack},
  howpublished={Spiceworks},
  year      = {2024},
  month     = {Feb.},
  @day       = {1},
  url       = {https://www.spiceworks.com/tech/artificial-intelligence/news/chatgpt-leaks-sensitive-user-data-openai-suspects-hack/},
urldate={2025-03-26},
  @note      = {Accessed: Mar. 26, 2025}
}

@misc{jin2024jailbreakzoo,
  title={{JailbreakZoo}: Survey, landscapes, and horizons in jailbreaking large language and vision-language models},
  author={Jin, Haibo and Hu, Leyang and Li, Xinuo and Zhang, Peiyan and Chen, Chonghan and Zhuang, Jun and Wang, Haohan},
  @note={\emph{arXiv:2407.01599}},
  year={2024},
      eprint={2407.01599},
      archivePrefix={arXiv},
      primaryClass={cs.CL},
}

@misc{hu2024firm,
  title={As Firm As Their Foundations: Can open-sourced foundation models be used to create adversarial examples for downstream tasks?},
  author={Hu, Anjun and Gu, Jindong and Pinto, Francesco and Kamnitsas, Konstantinos and Torr, Philip},
  @note={\emph{arXiv:2403.12693}},
  year={2024},
      eprint={2403.12693},
      archivePrefix={arXiv},
      primaryClass={cs.CV},
}

@misc{chen2023can,
  title={Can {LLM}-generated misinformation be detected?},
  author={Chen, Canyu and Shu, Kai},
  @note={\emph{arXiv:2309.13788}},
  year={2023},
      eprint={2309.13788},
      archivePrefix={arXiv},
      primaryClass={cs.CL},
}

@inproceedings{guo2024cold,
author = {Guo, Xingang and Yu, Fangxu and Zhang, Huan and Qin, Lianhui and Hu, Bin},
title = {{COLD}-attack: jailbreaking {LLMs} with stealthiness and controllability},
year = {2024},
publisher = {JMLR.org},
booktitle = {Proceedings of the 41st International Conference on Machine Learning},
articleno = {675},
numpages = {29},
location = {Vienna, Austria},
series = {ICML'24},
}

@inproceedings{mazeika2024harmbenchstandardizedevaluationframework,
author = {Mazeika, Mantas and Phan, Long and Yin, Xuwang and Zou, Andy and Wang, Zifan and Mu, Norman and Sakhaee, Elham and Li, Nathaniel and Basart, Steven and Li, Bo and Forsyth, David and Hendrycks, Dan},
title = {{HarmBench}: a standardized evaluation framework for automated red teaming and robust refusal},
year = {2024},
publisher = {JMLR.org},
booktitle = {Proceedings of the 41st International Conference on Machine Learning},
articleno = {1431},
numpages = {44},
location = {Vienna, Austria},
series = {ICML'24}
}

@inproceedings{zhou2024robustpromptoptimizationdefending,
 author = {Zhou, Andy and Li, Bo and Wang, Haohan},
 booktitle = {Advances in Neural Information Processing Systems},
 editor = {A. Globerson and L. Mackey and D. Belgrave and A. Fan and U. Paquet and J. Tomczak and C. Zhang},
 pages = {40184--40211},
 publisher = {Curran Associates, Inc.},
 title = {Robust Prompt Optimization for Defending Language Models Against Jailbreaking Attacks},
 @url = {https://proceedings.neurips.cc/paper_files/paper/2024/file/46ed503889ab232c21c1162340ee17b2-Paper-Conference.pdf},
 volume = {37},
 year = {2024}
}

@article{zhu2023promptbench,
  author  = {Kaijie Zhu and Qinlin Zhao and Hao Chen and Jindong Wang and Xing Xie},
  title   = {PromptBench: A Unified Library for Evaluation of Large Language Models},
  journal = {Journal of Machine Learning Research},
  year    = {2024},
  volume  = {25},
  number  = {254},
  pages   = {1--22},
  @url     = {http://jmlr.org/papers/v25/24-0023.html}
}

@misc{li2024privacylargelanguagemodels,
      title={Privacy in Large Language Models: Attacks, Defenses and Future Directions}, 
      author={Haoran Li and Yulin Chen and Jinglong Luo and Jiecong Wang and Hao Peng and Yan Kang and Xiaojin Zhang and Qi Hu and Chunkit Chan and Zenglin Xu and Bryan Hooi and Yangqiu Song},
      year={2024},
      eprint={2310.10383},
      archivePrefix={arXiv},
      primaryClass={cs.CL},
@note={\emph{arXiv:2310.10383}},
      @url={https://arxiv.org/abs/2310.10383}, 
}

@inproceedings {papernot2019TensorFlow,
author = {Nicolas Papernot},
title = {Machine Learning at Scale with Differential Privacy in {TensorFlow}},
booktitle = {2019 {USENIX} Conference on Privacy Engineering Practice and Respect ({PEPR} 19)},
year = {2019},
address = {Santa Clara, CA},
url = {https://www.usenix.org/node/238163},
publisher = {USENIX Association},
month = aug
}

@inproceedings{feffer2024red,
  title={Red-Teaming for generative AI: Silver bullet or security theater?},
  author={Feffer, Michael and Sinha, Anusha and Deng, Wesley H and Lipton, Zachary C and Heidari, Hoda},
  booktitle={Proceedings of the AAAI/ACM Conference on AI, Ethics, and Society},
  volume={7},
  pages={421--437},
  year={2024},
doi={10.1609/aies.v7i1.31647},
@note = {doi: \href{http://dx.doi.org/10.1109/10.1609/aies.v7i1.31647}{10.1609/aies.v7i1.31647}},
}

@misc{derczynski2024garak,
  title={{garak}: A framework for security probing large language models},
  author={Derczynski, Leon and Galinkin, Erick and Martin, Jeffrey and Majumdar, Subho and Inie, Nanna},
  @note={\emph{arXiv:2406.11036}},
  year={2024},
      eprint={2406.11036},
      archivePrefix={arXiv},
      primaryClass={cs.CL},
}

@techreport{schultz2025deepseek,
  title        = {DeepSeek-r1 vs. {OpenAI}-o1: The Ultimate Security Showdown},
  author       = {Schultz, Dorian and Bilić, Ante and Jurinčić, Dominik and Granoša, Dorian and Kamber, Luka},
  institution  = {Splx.ai},
  year         = {2025},
  month        = {January},
  @note         = {Available from Splx.ai},
  url          = {https://splx.ai/blog/deepseek-r1-vs-openai-o1-the-ultimate-security-showdown},
}

@misc{wired2025deepseek,
  title   = {DeepSeek's Safety Guardrails Failed Every Test Researchers Threw at Its AI Chatbot},
  author  = {{WIRED}},
  year    = {2025},
  month   = {Jan.},
  url     = {https://www.wired.com/story/deepseeks-ai-jailbreak-prompt-injection-attacks/},
urldate = {2025-03-28},
  @note    = {{A}ccessed: Mar. 28, 2025}
}

@inproceedings{carlini2023extracting,
  title={Extracting training data from diffusion models},
  author={Carlini, Nicolas and Hayes, Jamie and Nasr, Milad and Jagielski, Matthew and Sehwag, Vikash and Tramer, Florian and Balle, Borja and Ippolito, Daphne and Wallace, Eric},
  booktitle={32nd USENIX Security Symposium (USENIX Security 23)},
  pages={5253--5270},
  year={2023}
}

@misc{perez2022ignorepreviouspromptattack,
      title={Ignore Previous Prompt: Attack Techniques For Language Models}, 
      author={Fábio Perez and Ian Ribeiro},
      year={2022},
      eprint={2211.09527},
      archivePrefix={arXiv},
      primaryClass={cs.CL},
@note={\emph{arXiv:2211.09527}},
      @url={https://arxiv.org/abs/2211.09527}, 
}

@book{hubbard2023cybersecurity,
  title     = {How to Measure Anything in Cybersecurity Risk},
  author    = {Douglas W. Hubbard and Richard Seiersen},
  edition   = {2nd},
  year      = {2023},
  publisher = {Wiley},
  isbn      = {978-1394163795},
doi={10.1002/9781119892335},
@note = {doi: \href{http://dx.doi.org/10.1002/9781119892335}{10.1002/9781119892335}},
}

@techreport{ibm2024breach,
  title        = {Cost of a Data Breach Report 2024},
  author       = {{IBM Security} and {Ponemon Institute}},
  year         = {2024},
  institution  = {IBM Corporation},
  @note      = {Accessed: Mar. 26, 2025},
  url = {https://www.ibm.com/reports/data-breach}}

@misc{moulton2025dsu,
  author       = {Moulton, Richard},
  title        = {{DSU}: QUANTIFYING RETURN ON CONTROLS IN {LLM} CYBERSECURITY},
  year         = {2025},
  url = {https://github.com/rhmoult/DSU},
  @note      = {Accessed: May 8, 2025},
  urldate      = {2025-05-08}
}

@inproceedings{brown2020language,
author = {Brown, Tom B. and Mann, Benjamin and Ryder, Nick and Subbiah, Melanie and Kaplan, Jared and Dhariwal, Prafulla and Neelakantan, Arvind and Shyam, Pranav and Sastry, Girish and Askell, Amanda and Agarwal, Sandhini and Herbert-Voss, Ariel and Krueger, Gretchen and Henighan, Tom and Child, Rewon and Ramesh, Aditya and Ziegler, Daniel M. and Wu, Jeffrey and Winter, Clemens and Hesse, Christopher and Chen, Mark and Sigler, Eric and Litwin, Mateusz and Gray, Scott and Chess, Benjamin and Clark, Jack and Berner, Christopher and McCandlish, Sam and Radford, Alec and Sutskever, Ilya and Amodei, Dario},
title = {Language models are few-shot learners},
year = {2020},
isbn = {9781713829546},
publisher = {Curran Associates Inc.},
booktitle = {Proceedings of the 34th International Conference on Neural Information Processing Systems (NIPS '20)},
articleno = {159},
numpages = {25},
location = {Vancouver, BC, Canada},
url = {https://proceedings.neurips.cc/paper_files/paper/2020/file/1457c0d6bfcb4967418bfb8ac142f64a-Paper.pdf}
}

@misc{bommasani2022opportunities,
      title={On the Opportunities and Risks of Foundation Models}, 
      author={Rishi Bommasani and Drew A. Hudson and Ehsan Adeli and Russ Altman and Simran Arora and Sydney von Arx and Michael S. Bernstein and Jeannette Bohg and Antoine Bosselut and Emma Brunskill and Erik Brynjolfsson and Shyamal Buch and Dallas Card and Rodrigo Castellon and Niladri Chatterji and Annie Chen and Kathleen Creel and Jared Quincy Davis and Dora Demszky and Chris Donahue and Moussa Doumbouya and Esin Durmus and Stefano Ermon and John Etchemendy and Kawin Ethayarajh and Li Fei-Fei and Chelsea Finn and Trevor Gale and Lauren Gillespie and Karan Goel and Noah Goodman and Shelby Grossman and Neel Guha and Tatsunori Hashimoto and Peter Henderson and John Hewitt and Daniel E. Ho and Jenny Hong and Kyle Hsu and Jing Huang and Thomas Icard and Saahil Jain and Dan Jurafsky and Pratyusha Kalluri and Siddharth Karamcheti and Geoff Keeling and Fereshte Khani and Omar Khattab and Pang Wei Koh and Mark Krass and Ranjay Krishna and Rohith Kuditipudi and Ananya Kumar and Faisal Ladhak and Mina Lee and Tony Lee and Jure Leskovec and Isabelle Levent and Xiang Lisa Li and Xuechen Li and Tengyu Ma and Ali Malik and Christopher D. Manning and Suvir Mirchandani and Eric Mitchell and Zanele Munyikwa and Suraj Nair and Avanika Narayan and Deepak Narayanan and Ben Newman and Allen Nie and Juan Carlos Niebles and Hamed Nilforoshan and Julian Nyarko and Giray Ogut and Laurel Orr and Isabel Papadimitriou and Joon Sung Park and Chris Piech and Eva Portelance and Christopher Potts and Aditi Raghunathan and Rob Reich and Hongyu Ren and Frieda Rong and Yusuf Roohani and Camilo Ruiz and Jack Ryan and Christopher Ré and Dorsa Sadigh and Shiori Sagawa and Keshav Santhanam and Andy Shih and Krishnan Srinivasan and Alex Tamkin and Rohan Taori and Armin W. Thomas and Florian Tramèr and Rose E. Wang and William Wang and Bohan Wu and Jiajun Wu and Yuhuai Wu and Sang Michael Xie and Michihiro Yasunaga and Jiaxuan You and Matei Zaharia and Michael Zhang and Tianyi Zhang and Xikun Zhang and Yuhui Zhang and Lucia Zheng and Kaitlyn Zhou and Percy Liang},
      year={2022},
      eprint={2108.07258},
      archivePrefix={arXiv},
      primaryClass={cs.LG},
      @url={https://arxiv.org/abs/2108.07258}, 
@note={\emph{arXiv:2108.07258}},
}

@misc{chen2021evaluatingllms,
      title={Evaluating Large Language Models Trained on Code}, 
      author={Mark Chen and Jerry Tworek and Heewoo Jun and Qiming Yuan and Henrique Ponde de Oliveira Pinto and Jared Kaplan and Harri Edwards and Yuri Burda and Nicholas Joseph and Greg Brockman and Alex Ray and Raul Puri and Gretchen Krueger and Michael Petrov and Heidy Khlaaf and Girish Sastry and Pamela Mishkin and Brooke Chan and Scott Gray and Nick Ryder and Mikhail Pavlov and Alethea Power and Lukasz Kaiser and Mohammad Bavarian and Clemens Winter and Philippe Tillet and Felipe Petroski Such and Dave Cummings and Matthias Plappert and Fotios Chantzis and Elizabeth Barnes and Ariel Herbert-Voss and William Hebgen Guss and Alex Nichol and Alex Paino and Nikolas Tezak and Jie Tang and Igor Babuschkin and Suchir Balaji and Shantanu Jain and William Saunders and Christopher Hesse and Andrew N. Carr and Jan Leike and Josh Achiam and Vedant Misra and Evan Morikawa and Alec Radford and Matthew Knight and Miles Brundage and Mira Murati and Katie Mayer and Peter Welinder and Bob McGrew and Dario Amodei and Sam McCandlish and Ilya Sutskever and Wojciech Zaremba},
      year={2021},
      eprint={2107.03374},
      archivePrefix={arXiv},
      primaryClass={cs.LG},
      @url={https://arxiv.org/abs/2107.03374}, 
@note={\emph{arXiv:2107.03374}},
}

@article{lai2024llmlaw,
title = {Large language models in law: A survey},
journal = {AI Open},
volume = {5},
pages = {181-196},
year = {2024},
issn = {2666-6510},
doi = {https://doi.org/10.1016/j.aiopen.2024.09.002},
@url = {https://www.sciencedirect.com/science/article/pii/S2666651024000172},
author = {Jinqi Lai and Wensheng Gan and Jiayang Wu and Zhenlian Qi and Philip S. Yu},
@note = {doi: \href{http://dx.doi.org/10.1016/j.aiopen.2024.09.002}{10.1016/j.aiopen.2024.09.002}},
}

@misc{bubeck2023sparks,
      title={Sparks of Artificial General Intelligence: Early experiments with GPT-4}, 
      author={Sébastien Bubeck and Varun Chandrasekaran and Ronen Eldan and Johannes Gehrke and Eric Horvitz and Ece Kamar and Peter Lee and Yin Tat Lee and Yuanzhi Li and Scott Lundberg and Harsha Nori and Hamid Palangi and Marco Tulio Ribeiro and Yi Zhang},
      year={2023},
      eprint={2303.12712},
      archivePrefix={arXiv},
      primaryClass={cs.CL},
      @url={https://arxiv.org/abs/2303.12712}, 
@note={\emph{arXiv:2303.12712}},
}

@techreport{hu2013guide,
  title={Guide to attribute based access control ({ABAC}) definition and considerations},
  author={Hu, Vincent C and Ferraiolo, David and Kuhn, Rick and Friedman, Arthur R and Lang, Alan J and Cogdell, Margaret M and Schnitzer, Adam and Sandlin, Kenneth and Miller, Robert and Scarfone, Karen and others},
  journal={NIST special publication},
  number={NIST Special Publication 800-162},
  pages={1--54},
  year={2013},
doi={10.6028/NIST.SP.800-162},
@note = {doi: \href{http://dx.doi.org/10.6028/NIST.SP.800-162}{10.6028/NIST.SP.800-162}},
}

@inproceedings{abadi2016deep,
author = {Abadi, Martin and Chu, Andy and Goodfellow, Ian and McMahan, H. Brendan and Mironov, Ilya and Talwar, Kunal and Zhang, Li},
title = {Deep Learning with Differential Privacy},
year = {2016},
isbn = {9781450341394},
publisher = {Association for Computing Machinery},
address = {New York, NY, USA},
@url = {https://doi.org/10.1145/2976749.2978318},
doi = {10.1145/2976749.2978318},
@note = {doi: \href{http://dx.doi.org/10.1145/2976749.2978318}{10.1145/2976749.2978318}},
booktitle = {Proceedings of the 2016 ACM SIGSAC Conference on Computer and Communications Security (CCS '16)},
pages = {308–318},
numpages = {11},
location = {Vienna, Austria},
}

@inproceedings{wenzek2020ccnet,
    title = "{CCN}et: Extracting High Quality Monolingual Datasets from Web Crawl Data",
    author = "Wenzek, Guillaume  and
      Lachaux, Marie-Anne  and
      Conneau, Alexis  and
      Chaudhary, Vishrav  and
      Guzm{\'a}n, Francisco  and
      Joulin, Armand  and
      Grave, Edouard",
    booktitle = "Proceedings of the Twelfth Language Resources and Evaluation Conference",
    month = may,
    year = "2020",
    @address = "Marseille, France",
    publisher = "European Language Resources Association",
    url = "https://aclanthology.org/2020.lrec-1.494/",
    pages = "4003--4012",
    ISBN = "979-10-95546-34-4",
}

@inproceedings{miyato2017adversarial,
title={Adversarial Training Methods for Semi-Supervised Text Classification},
author={Takeru Miyato and Andrew M. Dai and Ian Goodfellow},
booktitle={International Conference on Learning Representations},
year={2017}
}

@inproceedings{gehman2020real,
    title = "{R}eal{T}oxicity{P}rompts: Evaluating Neural Toxic Degeneration in Language Models",
    author = "Gehman, Samuel  and
      Gururangan, Suchin  and
      Sap, Maarten  and
      Choi, Yejin  and
      Smith, Noah A.",
    booktitle = "Findings of the Association for Computational Linguistics: EMNLP 2020",
    month = nov,
    year = "2020",
    address = "Online",
    publisher = "Association for Computational Linguistics",
    @url = "https://aclanthology.org/2020.findings-emnlp.301/",
    doi = "10.18653/v1/2020.findings-emnlp.301",
    pages = "3356--3369",
@note = {doi: \href{http://dx.doi.org/10.18653/v1/2020.findings-emnlp.301}{10.18653/v1/2020.findings-emnlp.301}},
}

@article{liang2023holistic,
title={Holistic Evaluation of Language Models},
author={Percy Liang and Rishi Bommasani and Tony Lee and Dimitris Tsipras and Dilara Soylu and Michihiro Yasunaga and Yian Zhang and Deepak Narayanan and Yuhuai Wu and Ananya Kumar and Benjamin Newman and Binhang Yuan and Bobby Yan and Ce Zhang and Christian Cosgrove and Christopher D Manning and Christopher Re and Diana Acosta-Navas and Drew A. Hudson and Eric Zelikman and Esin Durmus and Faisal Ladhak and Frieda Rong and Hongyu Ren and Huaxiu Yao and Jue WANG and Keshav Santhanam and Laurel Orr and Lucia Zheng and Mert Yuksekgonul and Mirac Suzgun and Nathan Kim and Neel Guha and Niladri S. Chatterji and Omar Khattab and Peter Henderson and Qian Huang and Ryan Andrew Chi and Sang Michael Xie and Shibani Santurkar and Surya Ganguli and Tatsunori Hashimoto and Thomas Icard and Tianyi Zhang and Vishrav Chaudhary and William Wang and Xuechen Li and Yifan Mai and Yuhui Zhang and Yuta Koreeda},
journal={Transactions on Machine Learning Research},
issn={2835-8856},
year={2023},
url={https://openreview.net/forum?id=iO4LZibEqW},
}

@inproceedings{zhang2025intention,
    title = "Intention Analysis Makes {LLM}s A Good Jailbreak Defender",
    author = "Zhang, Yuqi  and
      Ding, Liang  and
      Zhang, Lefei  and
      Tao, Dacheng",
    booktitle = "Proceedings of the 31st International Conference on Computational Linguistics",
    month = jan,
    year = "2025",
    address = "Abu Dhabi, UAE",
    publisher = "Association for Computational Linguistics",
    url = "https://aclanthology.org/2025.coling-main.199/",
    pages = "2947--2968",
}

@inproceedings{jin2020bert, 
title={Is {BERT} Really Robust? A Strong Baseline for Natural Language Attack on Text Classification and Entailment}, 
volume={34}, 
@url={https://ojs.aaai.org/index.php/AAAI/article/view/6311}, 
DOI={10.1609/aaai.v34i05.6311}, 
number={05}, 
booktitle={Proceedings of the AAAI Conference on Artificial Intelligence}, 
author={Jin, Di and Jin, Zhijing and Zhou, Joey Tianyi and Szolovits, Peter}, year={2020}, month={Apr.}, pages={8018-8025},
@note = {doi: \href{http://dx.doi.org/10.1609/aaai.v34i05.6311}{10.1609/aaai.v34i05.6311}}
}

@inproceedings{wallace2019universal,
    title = "Universal Adversarial Triggers for Attacking and Analyzing {NLP}",
    author = "Wallace, Eric  and
      Feng, Shi  and
      Kandpal, Nikhil  and
      Gardner, Matt  and
      Singh, Sameer",
    booktitle = "Proceedings of the 2019 Conference on Empirical Methods in Natural Language Processing and the 9th International Joint Conference on Natural Language Processing (EMNLP-IJCNLP)",
    month = nov,
    year = "2019",
    @address = "Hong Kong, China",
    publisher = "Association for Computational Linguistics",
    @url = "https://aclanthology.org/D19-1221/",
    doi = "10.18653/v1/D19-1221",
    pages = "2153--2162",
@note = {doi: \href{http://dx.doi.org/10.18653/v1/D19-1221}{10.18653/v1/D19-1221}}
}

@article{vidgen2020directions,
  title={Directions in abusive language training data, a systematic review: Garbage in, garbage out},
  author={Vidgen, Bertie and Derczynski, Leon},
  journal={PLOS one},
  volume={15},
  number={12},
  @pages={e0243300},
  year={2020},
doi={10.1371/journal.pone.0243300},
@note = {doi: \href{http://dx.doi.org/10.1371/journal.pone.0243300}{10.1371/journal.pone.0243300}}
}

@misc{EU_HLEG_TrustworthyAI_2019,
  author       = {{European Commission High-Level Expert Group on Artificial Intelligence}},
  title        = {Ethics Guidelines for Trustworthy {AI}},
  url = {https://digital-strategy.ec.europa.eu/en/library/ethics-guidelines-trustworthy-ai},
  year         = {2019},
  month        = {April},
  urldate         = {2025-08-11}
}

@article{jobin2019global,
  title={The global landscape of AI ethics guidelines},
  author={Jobin, Anna and Ienca, Marcello and Vayena, Effy},
  journal={Nature machine intelligence},
  volume={1},
  number={9},
  pages={389--399},
  year={2019},
  publisher={Nature Publishing Group UK London},
doi={10.1038/s42256-019-0088-2},
@note = {doi: \href{http://dx.doi.org/10.1038/s42256-019-0088-2}{10.1038/s42256-019-0088-2}}
}

\clearpage
\appendices
\onecolumn
\section{Loss Exceedance Curves}\label{AppA}


\clearpage
\noindent\mbox{}\\[2ex]  


\end{document}